\documentclass[twocolumn]{aastex63}
\def\actaa{Acta Astronomica}

\usepackage{amsmath}

\hypersetup{linkcolor=red,citecolor=blue,filecolor=green,urlcolor=magenta}

\begin{document}

\shorttitle{Type II Cepheid PL \& PW relations}
\shortauthors{Ngeow et al.}

\title{Zwicky Transient Facility and Globular Clusters: The Period-Luminosity and Period-Wesenheit Relations for Type II Cepheids}

\correspondingauthor{C.-C. Ngeow}
\email{cngeow@astro.ncu.edu.tw}

\author[0000-0001-8771-7554]{Chow-Choong Ngeow}
\affil{Graduate Institute of Astronomy, National Central University, 300 Jhongda Road, 32001 Jhongli, Taiwan}

\author[0000-0001-6147-3360]{Anupam Bhardwaj}
\affil{INAF-Osservatorio astronomico di Capodimonte, Via Moiariello 16, 80131 Napoli, Italy}

\author{Jing-Yi Henderson}
\affil{Graduate Institute of Astronomy, National Central University, 300 Jhongda Road, 32001 Jhongli, Taiwan}

\author[0000-0002-3168-0139]{Matthew J. Graham}
\affiliation{Division of Physics, Mathematics, and Astronomy, California Institute of Technology, Pasadena, CA 91125, USA}

\author[0000-0003-2451-5482]{Russ R. Laher}
\affiliation{IPAC, California Institute of Technology, 1200 E. California Blvd, Pasadena, CA 91125, USA}

\author[0000-0002-7226-0659]{Michael S. Medford}
\affiliation{University of California, Berkeley, Department of Astronomy, Berkeley, CA 94720, USA}
\affiliation{Lawrence Berkeley National Laboratory, 1 Cyclotron Rd., Berkeley, CA 94720, USA}

\author[0000-0003-1227-3738]{Josiah Purdum}
\affiliation{Caltech Optical Observatories, California Institute of Technology, Pasadena, CA 91125, USA} 

\author[0000-0001-7648-4142]{Ben Rusholme}
\affiliation{IPAC, California Institute of Technology, 1200 E. California Blvd, Pasadena, CA 91125, USA}

\begin{abstract}
  We present the first $gri$-band period-luminosity (PL) and period-Wesenheit (PW) relations for 37 Type II Cepheids (hereafter TIIC) located in 18 globular clusters based on photometric data from the Zwicky Transient Facility. We also updated $BVIJHK$-band absolute magnitudes for 58 TIIC in 24 globular clusters using the latest homogeneous distances to the globular clusters. The slopes of $g/r/i$ and $B/V/I$ band PL relations are found to be statistically consistent when using the same sample of distance and reddening. We employed the calibration of $ri$-band PL/PW relations in globular clusters to estimate a distance to M31 based on a sample of $\sim 270$ TIIC from the PAndromeda project. The distance modulus to M31, obtained using calibrated $ri$-band PW relation, agrees well with the recent determination based on classical Cepheids. However, distance moduli derived using the calibrated $r$- and $i$-band PL relations are systematically smaller by $\sim0.2$~mag, suggesting there are possible additional systematic error on the PL relations. Finally, we also derive the period-color (PC) relations and for the first time the period-Q-index (PQ) relations, where the $Q$-index is reddening-free, for our sample of TIIC. The PC relations based on $(r-i)$ and near-infrared colors and the PQ relations are found to be relatively independent of the pulsation periods.
  
\end{abstract}


\section{Introduction}\label{sec1}

The evolved and low-mass Type II Cepheids \citep[hereafter TIIC; for a general review, see][]{welch2012} are one of the old population distance indicators. Similar to the young Type I or classical Cepheids, TIIC also exhibit a period-luminosity (PL, or the Leavitt Law) relation. However, TIIC are $\sim 2$~mag less luminous than the classical Cepheids. Nevertheless, TIIC are a few magnitudes more luminous, depending on the pulsation periods and filters, than the popular RR Lyrae -- another old population distance indicator. Therefore, TIIC are useful to probe a more distant stellar system (such as dwarf galaxies and elliptical galaxies) independent of RR Lyrae stars. The comprehensive reviews on TIIC as distance indicators can be found, for examples, in \citet{wallerstein2002}, \citet{sandage2006}, \citet{beaton2018}, \citet{bhardwaj2020}, and \citet{bhardwaj2022}.

Some of the earlier derivations of $BVI$-band, or a subset of these filters, PL relations for TIIC can be found, for examples, in \citet{demers1971}, \citet{nemec1994}, \citet{alcock1998}, and \citet{pritzl2003}. Other works on the optical PL relations included a color term \citep{breger1975,alcock1998} to derive the period-luminosity-color (PLC) relation, or using the Wesenheit index to derive the equivalent period-Wesenheit (PW) relation \citep{kubiak2003,matsunaga2011,groenewegen2017}. Recently, the optical band PL and PW relations were extended to the filters specific for the {\it Gaia} mission \citep{ripepi2019,ripepi2022}. In addition, \citet{groenewegen2017} have also derived the bolometric PL relation based on a combined sample of TIIC in Magellanic Clouds.

Compared to the optical PL relations, more studies have derived TIIC PL and PW relations in the near-infrared $JHK$ bands, or a subset of these filters, in the past two decades. These near-infrared PL/PW relations were derived using TIIC located in various stellar systems, including globular clusters \citep{matsunaga2006}, the Galactic Bulge \citep{groenewegen2008,bhardwaj2017a,braga2018}, the Large and/or Small Magellanic Cloud \citep{matsunaga2009,ciechanowska2010,matsunaga2011,ripepi2015,bhardwaj2017b,wiegorski2021}, and in nearby Milky Way field \citep{wiegorski2021}. Some of the derived $K$-band PL relations in the Galactic bulge also included an additional dependence on the Galactic longitude and latitude \citep{groenewegen2008,braga2018}.

To our knowledge, there is no $ugrizY$-band PL and PW relations available in the literature, which will be important in the era of Vera Rubin Observatory Legacy Survey of Space and Time \citep[LSST,][]{lsst2019}. Therefore, the goal of this work is to derive the $gri$-band PL and PW relations, by utilizing the time-series observations from the Zwicky Transient Facility \citep[ZTF,][]{bellm2017,bellm2019,dec20,gra19} project and archival data compiled in \citet[][because ZTF cannot observe the southern sky]{bhardwaj2022}, for TIIC located in the globular clusters. TIIC in globular clusters have been used to derive PL relations in the past. \citet{demers1974} derived the $V$-band PL relation based on 17 TIIC found in 4 globular clusters, while \citet{pritzl2003} derived the $BVI$-band PL relations using two globular clusters (NGC 6388 and NGC 6441) that host the most TIIC (for a total of 10 TIIC). Optical and near-infrared PL relations were also derived from a larger sample of TIIC in \citet[][with $\sim 40$ TIIC in 15 globular clusters]{nemec1994} and \citet[][with 46 TIIC in 26 globular clusters]{matsunaga2006}, respectively. Note that PL relations presented in \citet{matsunaga2006} were updated in \citet{braga2020} and \citet{bhardwaj2022}. 

Section \ref{sec2} describes the TIIC sample and their ZTF light curves data used in this work. In Section \ref{sec3}, we refined the pulsation periods and determined the mean magnitudes for our sample of TIIC. The derivations of the PL relations are presented in Section \ref{sec4}, as well as the multi-band relations (PW and period-color relations) in Section \ref{sec5}. We tested our derived PL/PW relations for a sample of M31 TIIC in Section \ref{sec6}, followed by conclusions of our work in Section \ref{sec7}.

\section{Sample and Data} \label{sec2}

\subsection{Selecting TIIC in Globular Clusters} \label{sec2.1}

We started the compilation of TIIC in globular clusters using the ``Updated Catalog of Variable Stars in Globular Clusters'' \citep[][hereafter Clement's Catalog]{clement2001,clement2017}, by selecting globular clusters that can be observed with ZTF ($\delta_{J2000} > -30^\circ$) and variable stars marked as ``CW'', ``CWA'', ``CWB'', ``RV'', or ``RVB'' in the Clement's Catalog.\footnote{Classifications of variable stars in the Clement's Catalog were based on the GCVS (General Catalog of Variable Stars) classification, available at \url{http://www.sai.msu.su/gcvs/gcvs/vartype.htm}. In brief, ``CW'' refers to W Virginis type, ``CWA'' and ``CWB'' are subtypes of ``CW'' with pulsation periods separated at 8~days. ``RV'' refers to the RV Tauri type, and ``RVB'' is subtype of ``RV'' which exhibits long-term periodic variations. Both W Virginis and RV Tauri are also subtypes of TIIC.} The known foreground or suspected foreground TIIC in the Clement's Catalog (marked with an ``f'' or ``f?''), however, were excluded. The preliminary list of TIIC were augmented with the catalogs presented in \citet{pritzl2003} and \citet{matsunaga2006}. We have also searched the literature for new TIIC, and updated equatorial coordinates, periods, and classifications of TIIC in our preliminary list. We identified five new, or re-classified, TIIC: V24 in M10 \citep{rozyczka2018}, V167 in M14 \citep{yepez2022}, V34 and ZK3 in M15 \citep{bhardwaj2021}, and V24 in M22 \citep{rozyczka2017}. Similarly, we rejected the TIIC that were re-classified as other types of variable stars in recent work, they included V1 in M10 \citep[identified as a semi-regular variable in][]{rozyczka2018}, V72 and V142 in M15 \citep[identified as a RR Lyrae and an anomalous Cepheid, respectively, in][]{bhardwaj2021}, V21 and V22 in M28 \citep[identified as a long-period variable and a RR Lyrae, respectively, in][]{prieto2012}, V8 in M79 \citep[identified as a semi-regular variable in][]{bond2016}, and V7 in M92 \citep[identified as an anomalous Cepheid in][]{osborn2012}. We also excluded S7 in M3 because the position of this variable star coincides with V254, a known RR Lyrae. All together, our preliminary list contains 50 TIIC located in 23 globular clusters.

\subsection{Extracting ZTF Light-Curves} \label{sec2.2}

ZTF is a wide-field synoptic survey on the northern sky observed in $gri$ filters. Combining the Samuel Oschin 48 inch Schmidt telescope (located at the Palomar Observatory) and a dedicated wide-field mosaic CCD camera, the field-of-view of ZTF can reach to $47$~squared degrees, while maintaining a pixel scale of $1.01\arcsec$/pixel. ZTF carries out three high-level surveys: the partner surveys, the public surveys, and the Caltech (California Institute of Technology) surveys. Imaging data from all of these high-level surveys were processed through a dedicated reduction pipeline \citep{mas19}, and the photometry were calibrated to the Pan-STARRS1 \citep[Panoramic Survey Telescope and Rapid Response System 1,][]{chambers2016,magnier2020} AB magnitude system. The preliminary list of TIIC sample were cross-matched to the PSF (point-spread function) catalogs, generated from the reduction pipeline, using an $1\arcsec$~search radius. The extracted $gri$-band (whenever available) light-curves for these TIIC were based on the ZTF Public Data Release 10 (DR10) data and partner surveys data until 2022 March 31. Out of the preliminary 50 TIIC sample, 48 of them have ZTF light-curves in at least two of the $gri$ filters (there are 1 and 11 TIIC without the $g$- and $i$-band light curves, respectively). The number of data points per light curve varies from 1 to $\sim1500$ for the extracted light curves, with medians of $158$, $504$, and $51$ in the $gri$-band, respectively. Two TIIC without ZTF light-curves are V1 and V2 in M19.

\begin{figure*}
  \plottwo{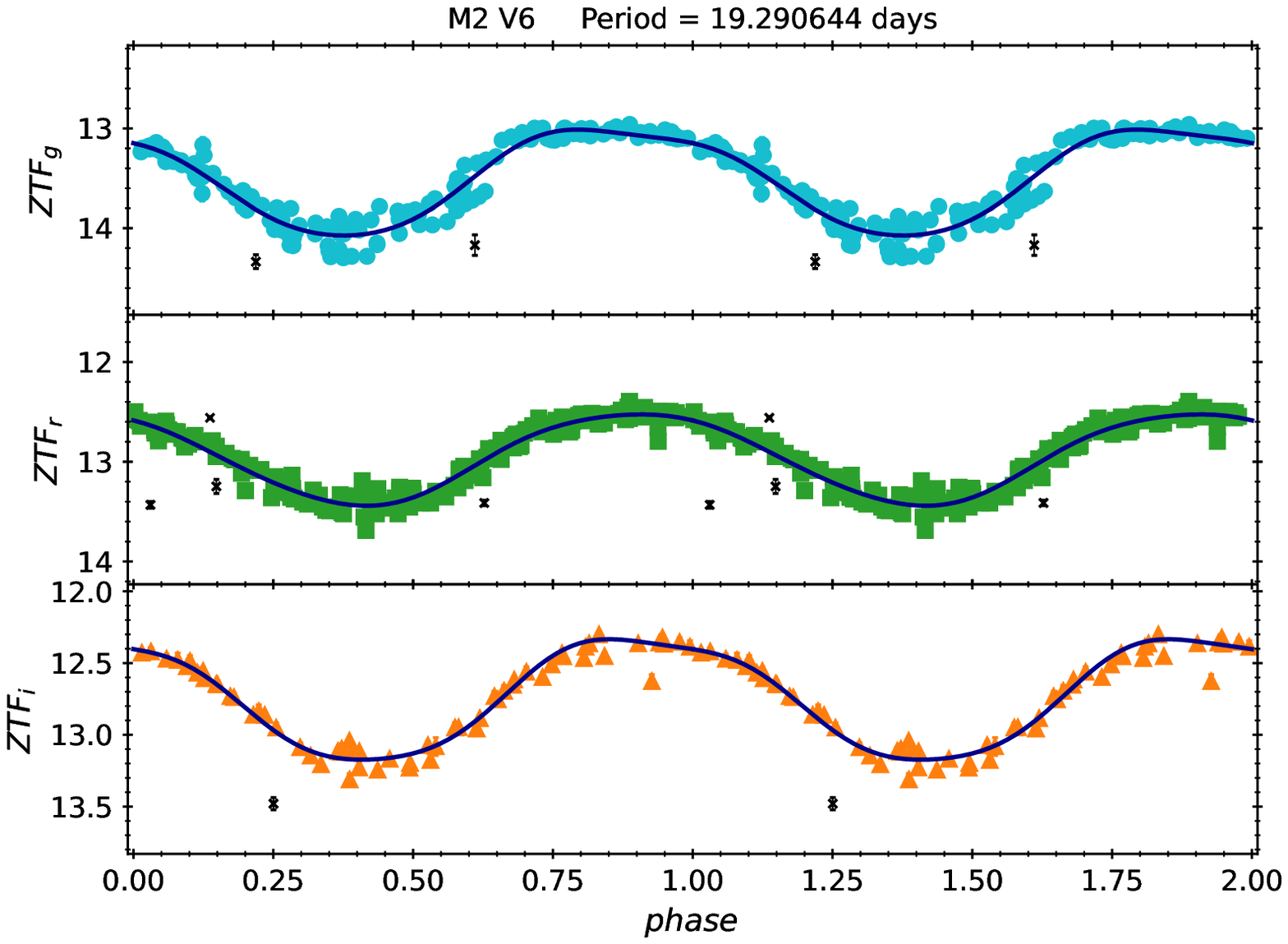}{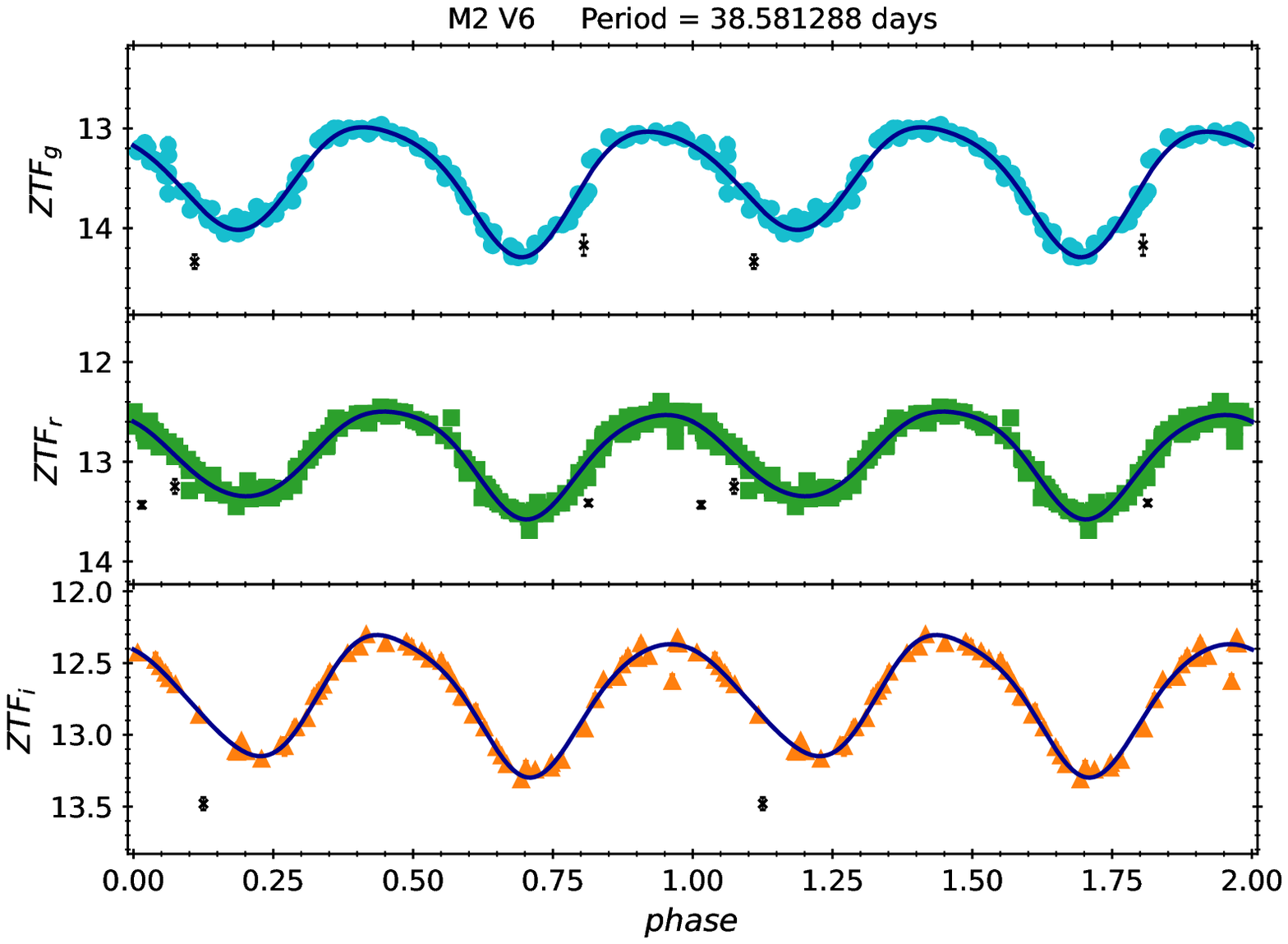}
  \caption{ZTF light-curves for V6 in M2 folded with the period determined from {\tt LombScargleMultiband} (left panel) and twice of the determined period (right panel). Alternate minima can be seen when the determined period is doubled. The black curves are fitted low-order Fourier expansion given in equation (1). Crosses are rejected outliers based on the two-steps fitting process (see text for details). }
  \label{fig_m2v6}
\end{figure*}

\begin{figure*}
  \gridline{\fig{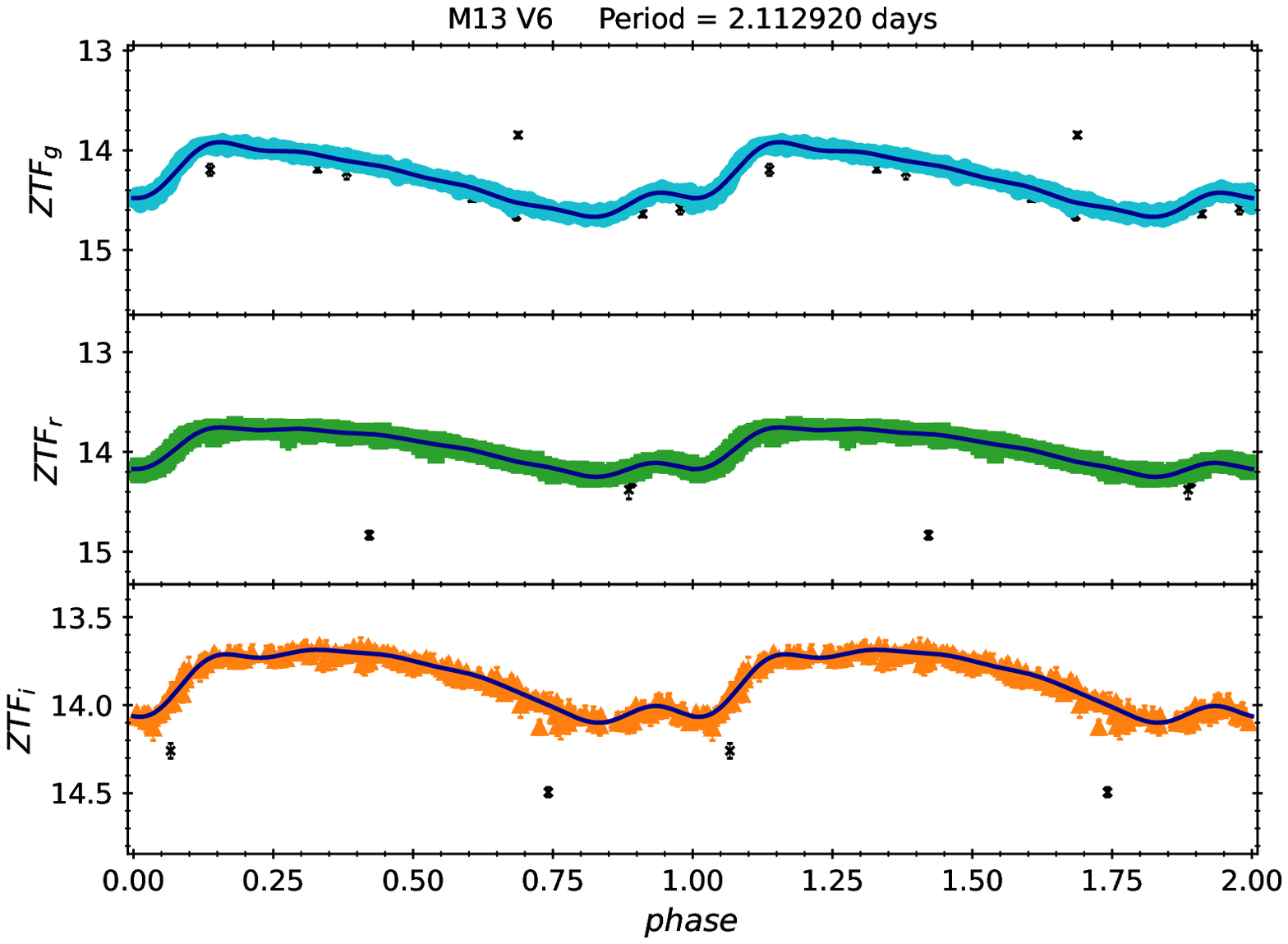}{0.32\textwidth}{BL Herculis sub-type}
    \fig{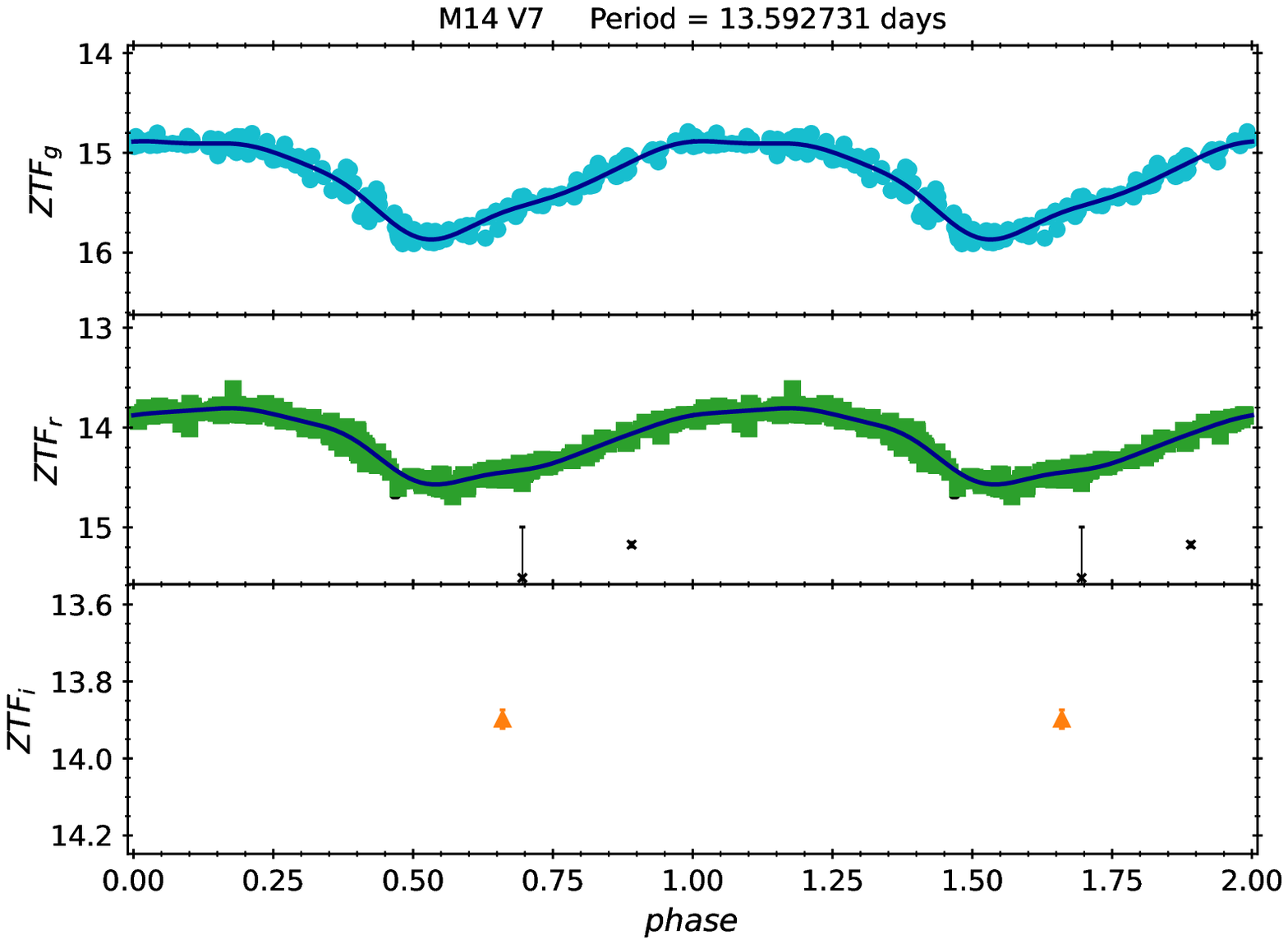}{0.32\textwidth}{W Virginis sub-type}
    \fig{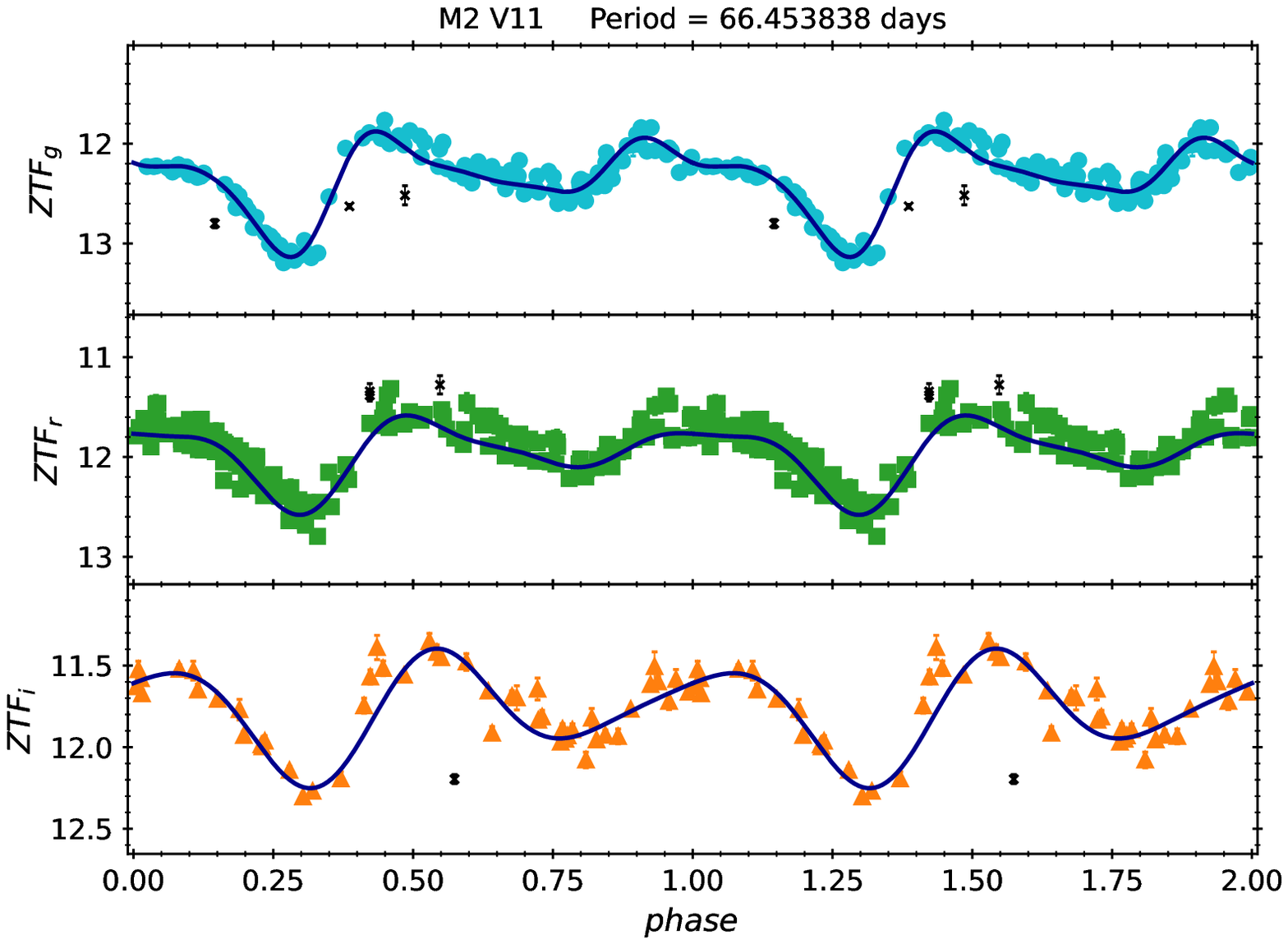}{0.32\textwidth}{RV Tauri sub-type}
  }
  \caption{Examples of ZTF light curves for TIIC in three different period ranges, roughly represent the three sub-types (BL Herculis, W Virginis, and RV Tauri) of TIIC. The black curves are fitted low-order Fourier expansion given in equation (1). Crosses are rejected outliers based on the two-steps fitting process (see text for details).}
  \label{fig_lc}
\end{figure*}

\section{Periods and Mean Magnitudes} \label{sec3}

Since it is well-known that TIIC will undergo periods change \citep[for examples, see][roughly in the range of $\sim10^{-8}$ to $\sim10^{-11}$~days/day]{wehlau1982,percy1997,percy2000,schmidt2004,schmidt2005a,schmidt2005b,rabidoux2010,osborn2012,soszynski2018,karmakar2019,berdnikov2021}, we re-determined the periods of our sample of TIIC with ZTF light curves instead of adopting the published periods.

\begin{deluxetable*}{llllrrrrrrrrl}
  \tabletypesize{\scriptsize}
  \tablecaption{Basic Information and Mean Magnitudes for ZTF Sample of TIIC in Globular Clusters\label{tab_t2cep}}
  \tablewidth{0pt}
  \tablehead{
    \colhead{G. C.} &
    \colhead{Var. Name} &
    \colhead{$P_{\mathrm{lit}}$\tablenotemark{a} (days)} &
    \colhead{$P$ (days)} &
    \colhead{$N_g$} &
    \colhead{$N_r$} &
    \colhead{$N_i$} &
    \colhead{$\langle g\rangle$} &
    \colhead{$\langle r\rangle$} &
    \colhead{$\langle i\rangle$} &
    \colhead{$D$\tablenotemark{b} (kpc)} &
    \colhead{$E$\tablenotemark{c}} &
    \colhead{Note\tablenotemark{d}}
  }
  \startdata
  M15	& V1	& 1.43781	& 1.437812	& 540	 & 665	 & 135	 & 15.029 & 14.837 & 14.754 & $10.71\pm0.10$ & $0.068\pm0.002$  & 1 \\
  M13	& V1	& 1.45902	& 1.459040	& 1067	 & 1103	 & 215	 & 14.173 & 14.037 & 14.011 & $7.42\pm0.08$ & $0.000\pm0.000$  & 3 \\
  M56	& V1	& 1.51000	& 1.509997	& 392	 & 851	 & 31	 & 15.685 & 15.232 & 15.049 & $10.43\pm0.14$ & $0.202\pm0.002$  & 8 \\
  NGC2419& V18	& 1.57870	& 1.578572	& 379	 & 1078	 & 63	 & 19.032 & 18.733 & 18.633 & $88.47\pm2.40$ & $0.144\pm0.004$  & 8 \\
  M22	& V11	& 1.69050	& 1.690401	& 77	 & 581	 & 0	 & 12.835 & 12.230 & $\cdots$ & $3.30\pm0.04$ & $0.419\pm0.006$  & 8 \\
  M22	& V24	& 1.71485	& 1.715079	& 76	 & 581	 & 0	 & 13.746 & 13.130 & $\cdots$ & $3.30\pm0.04$ & $0.419\pm0.006$  & 5 \\
  M15	& ZK3	& 1.74634	& 1.746591	& 537	 & 667	 & 134	 & 15.361 & 15.000 & 14.799 & $10.71\pm0.10$ & $0.162\pm0.004$  & 1 \\
  NGC6401& V3	& 1.74870	& 1.747028	& 83	 & 343	 & 71	 & 17.092 & 15.947 & 15.317 & $8.06\pm0.24$ & $0.926\pm0.002$  & 8 \\
  M14	& V76	& 1.88990	& 1.890065	& 182	 & 569	 & 1	 & 16.329 & 15.508 & $\cdots$ & $9.14\pm0.25$ & $0.540\pm0.000$  & 7 \\
  M13	& V6	& 2.11286	& 2.112920	& 1049	 & 1077	 & 215	 & 14.271 & 13.962 & 13.854 & $7.42\pm0.08$ & $0.000\pm0.000$  & 3 \\
  M10	& V24	& 2.30746	& 2.307591	& 71	 & 142	 & 1	 & 14.355 & 13.728 & $\cdots$ & $5.07\pm0.06$ & $0.312\pm0.002$  & 6 \\
  M19	& V4	& 2.43260	& 2.432354	& 62	 & 411	 & 0	 & 15.555 & 14.943 & $\cdots$ & $8.34\pm0.16$ & $0.488\pm0.005$  & 8 \\
  M14	& V2	& 2.79490	& 2.794852	& 182	 & 582	 & 1	 & 15.955 & 15.093 & $\cdots$ & $9.14\pm0.25$ & $0.540\pm0.000$  & 7 \\
  NGC6284& V4	& 2.81870	& 2.818707	& 63	 & 486	 & 0	 & 16.029 & 15.446 & $\cdots$ & $14.21\pm0.42$ & $0.318\pm0.002$  & 8 \\
  NGC6749& V1	& 4.48100	& 4.477411	& 125	 & 296	 & 2	 & 18.515 & 16.633 & $\cdots$ & $7.59\pm0.21$ & $1.346\pm0.007$  & 8 \\
  NGC6284& V1	& 4.48120	& 4.484024	& 66	 & 493	 & 0	 & 15.806 & 15.131 & $\cdots$ & $14.21\pm0.42$ & $0.318\pm0.002$  & 8 \\
  M13	& V2	& 5.11078	& 5.111326	& 1071	 & 1097	 & 216	 & 13.157 & 12.882 & 12.787 & $7.42\pm0.08$ & $0.000\pm0.000$  & 3 \\
  M14	& V167	& 6.20100	& 6.205786	& 182	 & 564	 & 1	 & 16.046 & 14.965 & $\cdots$ & $9.14\pm0.25$ & $0.560\pm0.003$  & 7 \\
  NGC6325& V2	& 10.74400	& 10.748907	& 66	 & 498	 & 0	 & 16.533 & 14.938 & $\cdots$ & $7.53\pm0.32$ & $0.966\pm0.005$  & 8 \\
  M14	& V17	& 12.07580	& 12.092216	& 184	 & 582	 & 1	 & 15.189 & 14.123 & $\cdots$ & $9.14\pm0.25$ & $0.540\pm0.000$  & 7 \\
  NGC6325& V1	& 12.51600	& 12.522716	& 65	 & 497	 & 0	 & 16.299 & 14.716 & $\cdots$ & $7.53\pm0.32$ & $0.928\pm0.006$  & 8 \\
  M28	& V4	& 13.46200	& 13.480377	& 136	 & 909	 & 144	 & 13.532 & 12.558 & 12.093 & $5.37\pm0.10$ & $0.458\pm0.004$  & 8 \\
  M14	& V7	& 13.58970	& 13.592731	& 185	 & 581	 & 1	 & 15.222 & 14.104 & $\cdots$ & $9.14\pm0.25$ & $0.560\pm0.003$  & 7 \\
  M79	& V7	& 13.99950	& 14.057529	& 114	 & 136	 & 0	 & 13.824 & 13.304 & $\cdots$ & $13.08\pm0.18$ & $0.014\pm0.002$  & 2 \\
  NGC6229& V8	& 14.84600	& 14.844260	& 1469	 & 1484	 & 431	 & 15.699 & 15.117 & 14.939 & $30.11\pm0.47$ & $0.092\pm0.002$  & 8 \\
  M2	& V1	& 15.56470	& 15.542598	& 61	 & 70	 & 6	 & 13.596 & 13.075 & $\cdots$ & $11.69\pm0.11$ & $0.000\pm0.000$  & 8 \\
  M80	& V1	& 16.28134	& 16.306309	& 62	 & 74	 & 0	 & 13.734 & 13.097 & $\cdots$ & $10.34\pm0.12$ & $0.220\pm0.003$  & 4 \\
  M19	& V3	& 16.50000	& 16.686135	& 66	 & 421	 & 0	 & 14.128 & 13.157 & $\cdots$ & $8.34\pm0.16$ & $0.488\pm0.005$  & 8 \\
  M15	& V86	& 16.84211	& 16.833319	& 514	 & 650	 & 133	 & 13.112 & 12.553 & 12.353 & $10.71\pm0.10$ & $0.162\pm0.004$  & 1 \\
  M2	& V5	& 17.55700	& 17.574309	& 132	 & 215	 & 53	 & 13.572 & 13.015 & 12.831 & $11.69\pm0.11$ & $0.004\pm0.004$  & 8 \\
  M10	& V2	& 19.47099	& 18.713201	& 146	 & 333	 & 2	 & 12.211 & 11.504 & $\cdots$ & $5.07\pm0.06$ & $0.312\pm0.002$  & 6 \\
  M14	& V1	& 19.74110	& 18.749399	& 184	 & 581	 & 1	 & 14.762 & 13.692 & $\cdots$ & $9.14\pm0.25$ & $0.568\pm0.002$  & 7 \\
  M5	& V42	& 25.73500	& 25.710120	& 199	 & 316	 & 75	 & 11.457 & 11.123 & 10.927 & $7.48\pm0.06$ & $0.090\pm0.000$  & 8 \\
  M2	& V6	& 19.29900	& 38.581288	& 156	 & 257	 & 61	 & 13.438 & 12.892 & 12.696 & $11.69\pm0.11$ & $0.000\pm0.000$  & 8 \\
  M5	& V84	& 53.95000	& 52.934619	& 245	 & 424	 & 100	 & 11.626 & 11.231 & 11.039 & $7.48\pm0.06$ & $0.112\pm0.002$  & 8 \\
  M2	& V11	& 67.00000	& 66.453838	& 132	 & 218	 & 50	 & 12.300 & 11.933 & 11.755 & $11.69\pm0.11$ & $0.000\pm0.000$  & 8 \\
  M56	& V6	& 90.00000	& 89.320054	& 391	 & 857	 & 31	 & 13.278 & 12.386 & 11.827 & $10.43\pm0.14$ & $0.202\pm0.002$  & 8 \\
  \enddata
  \tablenotetext{a}{Period published in the literature.}
  \tablenotetext{b}{Distance of the globular clusters adopted from \citet{baumgardt2021}.}
  \tablenotetext{c}{Reddening returned from the {\tt Bayerstar2019} 3D reddening map \citep{green2019} at the location of the TIIC with distance $D$ from \citet{baumgardt2021}.}
  \tablenotetext{d}{Literature period adopted from the following reference: 1 = \citet{bhardwaj2021}; 2 = \citet{bond2016}; 3 = \citet{osborn2019}; 4 = \citet{plachy2017}; 5 = \citep{rozyczka2017}; 6 = \citet{rozyczka2018}; 7 = \citet{yepez2022}; 8 = Clement's Catalog.} 
\end{deluxetable*}

Given that majority of our sample of TIIC have ZTF light curves in two or three filters, we employed the {\tt LombScargleMultiband} module available in the {\tt astroML/gatspy}\footnote{\url{https://github.com/astroML/gatspy}, also see \citet{vdp2016}.} package \citep{vdp2015} to refine the periods for our sample of TIIC in a two-steps process. In the first step, ZTF light-curves were folded using periods identified from the first-pass of {\tt LombScargleMultiband}, and then fit with a low-order Fourier expansion in the following form \cite[for example, see][]{deb2009}:

\begin{eqnarray}
  m(\Phi) & = & m_0 + \sum^n_{j=1} \left[ a_j \cos (2 \pi j  \Phi) + b_j \sin (2 \pi j  \Phi)\right],
\end{eqnarray}

\noindent where $\Phi \in [0,1]$ are the pulsational phases. Note that we only fit equation (1) to the light curves that have more than 30 data points. Outliers beyond $3\sigma$ were excluded, where $\sigma$ represents the dispersion of the fitted light curves, and {\tt LombScargleMultiband} was run again in the second-pass to obtain the final adopted periods. The periods obtained from {\tt LombScargleMultiband} need to be doubled for three TIIC (V11 in M2, V84 in M5, and V6 in M56) in order to match with published periods. We found that the period for V6 in M2 also needs to be doubled, because alternate minima can be seen on its light-curves (as displayed in Figure \ref{fig_m2v6}). 

\begin{figure*}
  \epsscale{1.1}
  \plottwo{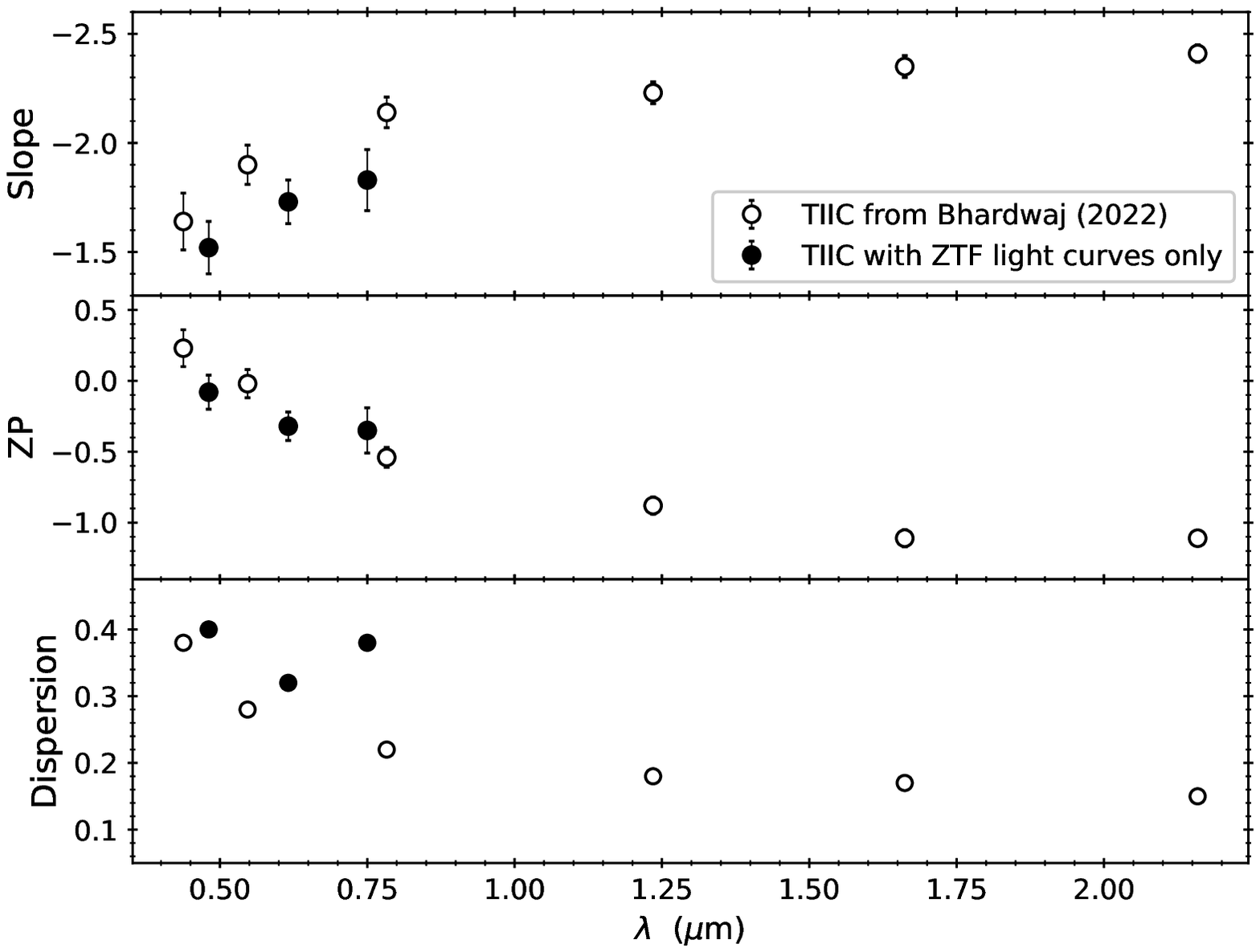}{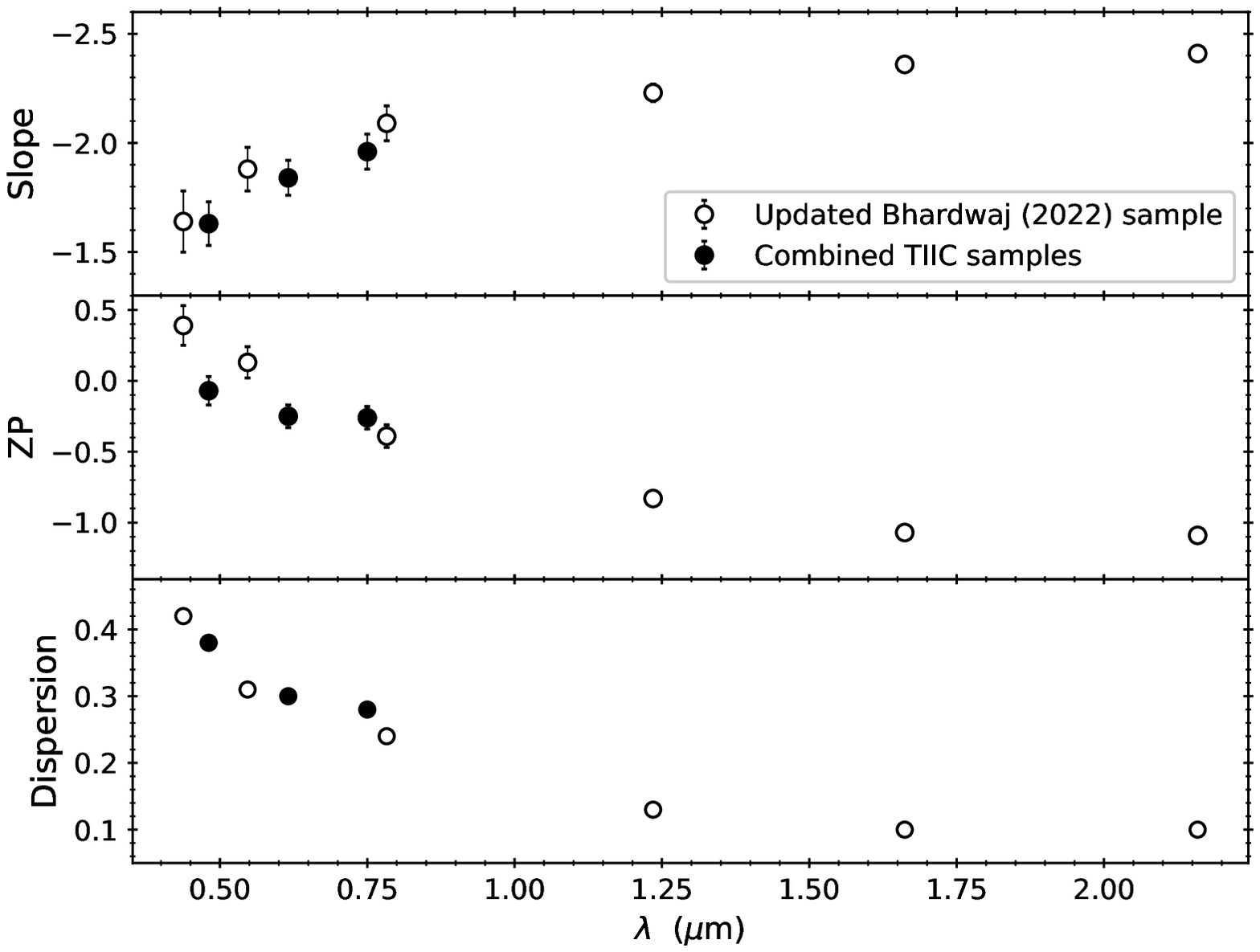}
  \caption{{\bf Left Panel:} Comparison of the slopes (upper panel), zero-points (ZP, middle panel), and the dispersions (lower panel) for PL relations derived in \citet[][for $BVIJHK$-band, in open symbols]{bhardwaj2022} and the preliminary $gri$-band PL relations using TIIC listed in Table \ref{tab_t2cep} (in filled symbols). {\bf Right Panel:} Same as the left panel, but for the updated PL relations as described in Section \ref{sec4.2}.}
  \label{fig_compare}
\end{figure*}

We visually inspected all light-curves folded with the final adopted periods. We removed 9 TIIC (V1 in M12, V12 in M13, V34 in M15, V17 and V32 in M28, V22 in NGC6229, V2 in NGC6293, V4 in NGC7492, and V4 in Pal3) from our sample because they exhibit evidence of blending (such as no variations or large scatters seen on the ZTF light-curves). We further removed 2 TIIC (V154 in M3 and V3 in M10) that only have 19 data points in the $r$-band light-curve (and the total number of data points in all three filters is 30 or less). Finally, 37 TIIC remained in our sample and their intensity mean magnitudes were obtained based on the fitted low-order Fourier expansion as given in equation (1). The final adopted periods and the intensity mean magnitudes of these TIIC are listed in Table \ref{tab_t2cep}. Examples of the ZTF light-curves are presented in Figure \ref{fig_lc}.

\section{The PL Relations} \label{sec4}

\subsection{Preliminary PL Relations} \label{sec4.1}

Homogeneous and accurate distances of globular clusters were adopted from \citet{baumgardt2021}, who combined various distance measurements based on the {\it Gaia} and/or {\it Hubble Space Telescope} data, as well as literature distances, to obtain averaged distances via a likelihood analysis. Using these distances, we queried the {\tt Bayerstar2019} 3D reddening map \citep{green2019}\footnote{See \url{http://argonaut.skymaps.info/usage}\label{fn2}} via the {\tt dustmaps}\footnote{\url{https://dustmaps.readthedocs.io/en/latest/}} \citep{green2018} code to obtain reddening $E$ towards each of the TIIC, and corrected the extinctions on mean magnitudes using $A_g = 3.518E$, $A_r = 2.617E$ and $A_i=1.971E$ \citep{green2019}. A linear regression was fitted to the extinction-corrected absolute magnitudes for 37 and 17 TIIC in the $gr$- and $i$-band, respectively. While fitting the PL relations, we did not separate the TIIC into the three sub-types (BL Herculis, W Virginis, and RV Tauri) of TIIC, mainly due to the small number of samples in each subtype.

We compare our preliminary $gri$-band PL relations to the Johnson-Cousin $BVI$-band and 2MASS $JHK$-band (hereafter collectively referred as $BVIJHK$-band) PL relations, taken from \citet{bhardwaj2022}, in the left panel of Figure \ref{fig_compare}. The slopes of the $gri$-band PL relations follow the trend that the slopes become steeper at longer wavelengths, however these $gri$-band PL slopes were shallower than the expected trends portrait from the $BVIJHK$-band PL slopes. Similar to our work, the $BVIJHK$-band PL relations were derived by \citet{bhardwaj2022} using a sample of 36 to 50 TIIC in globular clusters compiled from the literature. The distance moduli of these globular clusters were collected in \citet{braga2020}. In contrast to our work, these distance moduli were compiled from various publications \citep[see the reference listed in Table 4 of][]{braga2020}. In the next sub-section, we demonstrate that after updating the multi-band PL relations, the $gri$-band PL slopes are consistent with the $BVI$-band PL slopes, as shown in the upper-right panel of Figure \ref{fig_compare}. Similarly, the dispersion of the preliminary $gri$-band PL relations were larger (especially in the $i$-band), and improvements were evident after updating the PL relations.

\begin{deluxetable*}{lllllllllrcl}
  \tabletypesize{\scriptsize}
  \tablecaption{Basic Information and Mean Magnitudes for B22 Sample of TIIC in Globular Clusters\label{tab_b22}}
  \tablewidth{0pt}
  \tablehead{
    \colhead{G. C.} &
    \colhead{Var. Name} &
     \colhead{$P$ (days)} &
    \colhead{$B$} &
    \colhead{$V$} &
    \colhead{$I$} &
    \colhead{$J$} &
    \colhead{$H$} &
    \colhead{$K$} &
    \colhead{$D$\tablenotemark{a} (kpc)} &
    \colhead{$E(B-V)$\tablenotemark{b}} &
    \colhead{Reference\tablenotemark{c}}
  }
  \startdata
NGC5139  & V43  & 1.1569	& 14.139   & 13.759   & 13.149   & 12.730   & 12.492   & 12.426   & $5.43\pm0.05$ & 0.14  & 3 \\
NGC5139  & V92  & 1.346	        & 14.480   & 13.946   & 13.199   & 12.700   & 12.340   & 12.313   & $5.43\pm0.05$ & 0.13  & 3 \\
NGC5139  & V60  & 1.3495	& 14.028   & 13.624   & 13.001   & 12.584   & 12.295   & 12.281    & $5.43\pm0.05$ & 0.14  & 3  \\
M15      & V1   & 1.4377	& 15.412   & 14.954   & 14.362   & 13.94    & $\cdots$ & 13.65    & $10.71\pm0.10$ & 0.11  & 8 \\
M56      & V1   & 1.51	        & 16.01    & 15.46    & $\cdots$ & 13.99    & 13.66    & 13.57    & $10.43\pm0.14$ & 0.25  & 18 \\
M62      & V73  & 1.7	        & 16.147   & 15.243   & 13.966   & $\cdots$ & $\cdots$ & $\cdots$ & $6.41\pm0.10$ & 0.45  & 6, 17 \\
NGC2808  & V10  & 1.7653	& 15.91    & 15.28    & 14.47    & 13.89    & 13.54    & 13.43    & $10.06\pm0.11$ & 0.22  &  12 \\
M14      & V76  & 1.8903	& 16.881   & 15.978   & 14.750   & 13.78    & 13.30    & 13.16    & $9.14\pm0.25$ & 0.48  & 7 \\
M15      & V34  & 2.03355	& $\cdots$ & $\cdots$ & $\cdots$ & 13.756   & $\cdots$ & 13.340   & $10.71\pm0.10$ & 0.11  &  \\
NGC5139  & V61  & 2.2736	& 14.293   & 13.661   & 12.821   & 12.190   & 11.811   & 11.771   & $5.43\pm0.05$ & 0.14  & 3 \\
M19      & V4   & 2.4326	& 14.75    & $\cdots$ & 13.947   & 13.28    & 12.85    & 12.77    & $8.34\pm0.16$ & 0.31  & 4, 17  \\
NGC6441  & V132 & 2.5474	& 17.218   & 16.478   & 15.241   & $\cdots$ & $\cdots$ & $\cdots$ & $12.73\pm0.16$ & 0.61  &  13 \\
M14      & V2   & 2.7947	& 16.596   & 15.629   & 14.337   & 13.45    & 12.98    & 12.85    & $9.14\pm0.25$ & 0.48  & 7 \\
NGC6284  & V4   & 2.8187	& 16.04    & $\cdots$ & 14.786   & 14.15    & 13.71    & 13.67    & $14.21\pm0.42$ & 0.31  & 5, 17 \\
NGC5139  & V48  & 4.4752	& 13.528   & 12.924   & 12.092   & 11.59    & 11.14    & 11.15    & $5.43\pm0.05$ & 0.14  & 3 \\
NGC6749  & V1   & 4.481	        & $\cdots$ & $\cdots$ & $\cdots$ & 13.38    & 12.62    & 12.34    & $7.59\pm0.21$ & 1.75  &  \\
NGC6284  & V1   & 4.4812	& 15.88    & $\cdots$ & 14.504   & 13.68    & 13.24    & 13.18    & $14.21\pm0.42$ & 0.30  & 5, 17 \\
M10      & V3   & 7.831	        & 13.62    & 12.75    & 11.721   & 11.02    & 10.55    & 10.36    & $5.07\pm0.06$ & 0.27  & 2, 15 \\
NGC6441  & V153 & 9.89	        & $\cdots$ & $\cdots$ & 13.72    & $\cdots$ & $\cdots$ & $\cdots$ & $12.73\pm0.16$ & 0.62  & 16 \\
M62      & V2   & 10.59	        & 14.408   & 13.418   & 12.065   & 11.22    & 10.64    & 10.53    & $6.41\pm0.10$ & 0.47  & 6, 17 \\
NGC6325  & V2   & 10.744	& $\cdots$ & $\cdots$ & 13.632   & 12.14    & 11.43    & 11.22    & $7.53\pm0.32$ & 0.96  & 17 \\
NGC6441  & V154 & 10.83	        & $\cdots$ & $\cdots$ & 13.57    & $\cdots$ & $\cdots$ & $\cdots$ & $12.73\pm0.16$ & 0.61  & 16 \\
M14      & V17  & 12.091	& 15.846   & 14.676   & 13.182   & $\cdots$ & $\cdots$ & $\cdots$ & $9.14\pm0.25$ & 0.47  & 7 \\
NGC6256  & V1   & 12.447	& $\cdots$ & $\cdots$ & 13.402   & 11.86    & 11.15    & 10.85    & $7.24\pm0.29$ & 1.71  & 17 \\
NGC6325  & V1   & 12.516	& $\cdots$ & $\cdots$ & 13.436   & 11.97    & 11.25    & 11.02    & $7.53\pm0.32$ & 0.95  & 17 \\
M28      & V4   & 13.462	& 14.21    & $\cdots$ & 11.734   & 10.78    & 10.18    & 10.01    & $5.37\pm0.10$ & 0.49  & 17, 19 \\
NGC6441  & V128 & 13.519	& 16.475   & 15.257   & 13.795   & $\cdots$ & $\cdots$ & $\cdots$ & $12.73\pm0.16$ & 0.61  & 13 \\
M14      & V7   & 13.6038	& 16.051   & 14.745   & 13.224   & 12.04    & 11.46    & 11.29    & $9.14\pm0.25$ & 0.48  & 7 \\
M19      & V2   & 14.139	& 14.15    & $\cdots$ & 12.242   & 11.53    & 11.06    & 10.92    & $8.34\pm0.16$ & 0.32  & 4, 17 \\
HP1      & V17  & 14.42	        & $\cdots$ & $\cdots$ & $\cdots$ & 11.91    & 11.09    & 10.78    & $7.00\pm0.14$ & 2.32  &  \\
NGC5139  & V29  & 14.7338	& 12.776   & 12.015   & 11.049   & 10.43    & 10.03    & 9.93     & $5.43\pm0.05$ & 0.14  & 3 \\
M3       & V154 & 15.29	        & 12.79    & 12.33    & 11.68    & 11.45    & 11.06    & 10.99    & $10.18\pm0.08$ & 0.01  & 14 \\
M12      & V1   & 15.527	& $\cdots$ & $\cdots$ & $\cdots$ & 10.24    & 9.79     & 9.64     & $5.11\pm0.05$ & 0.18  &  \\
M2       & V1   & 15.5647	& 13.97    & 13.36    & $\cdots$ & 11.93    & 11.54    & 11.45    & $11.69\pm0.11$ & 0.04  & 9 \\
M80      & V1   & 16.3042	& 14.19    & 13.365   & $\cdots$ & 11.65    & 11.23    & 11.10    & $10.34\pm0.12$ & 0.21  & 11, 20 \\
HP1      & V16  & 16.4	        & $\cdots$ & $\cdots$ & $\cdots$ & 11.77    & 10.99    & 10.70    & $7.00\pm0.14$ & 2.39  &  \\
M19      & V3   & 16.5	        & 13.70    & $\cdots$ & 12.417   & $\cdots$ & $\cdots$ & $\cdots$ & $8.34\pm0.16$ & 0.31  & 4, 17 \\
M15      & V86  & 16.829	& 14.368   & 13.659   & 12.646   & 11.70    & 11.32    & 11.19    & $10.71\pm0.10$ & 0.11  & 8 \\
M19      & V1   & 16.92	        & 13.85    & $\cdots$ & 12.260   & 11.37    & 10.88    & 10.75    & $8.34\pm0.16$ & 0.32  & 4, 17 \\
M2       & V5   & 17.557	& 13.89    & 13.28    & $\cdots$ & 11.80    & 11.40     & 11.31    & $11.69\pm0.11$ & 0.04  & 9 \\
NGC6441  & V129 & 17.832	& 16.395   & 15.128   & 13.610   & 12.14    & 11.61    & 11.65    & $12.73\pm0.16$ & 0.62  & 13 \\
M10      & V2   & 18.7226	& 13.01    & 12.05   & 10.934    & 10.05    & 9.61     & 9.47     & $5.07\pm0.06$ & 0.29  &  2, 15 \\
M14      & V1   & 18.729	& 15.429   & 14.210   & 12.633   & 11.63    & 11.10    & 10.89    & $9.14\pm0.25$ & 0.48  & 7 \\
Terzan1  & V5   & 18.85	        & $\cdots$ & $\cdots$ & 14.576   & 11.97    & 10.93    & 10.61    & $5.67\pm0.17$ & 6.86  & 17 \\
M2       & V6   & 19.299	& 13.74    & 13.14    & $\cdots$ & 11.72    & 11.33    & 11.25    & $11.69\pm0.11$ & 0.04  & 9 \\
NGC6441  & V127 & 19.773	& 16.398   & 15.048   & 13.441   & $\cdots$ & $\cdots$ & $\cdots$ & $12.73\pm0.16$ & 0.61  & 13 \\
NGC6441  & V126 & 20.625	& 16.282   & 14.997   & 13.402   & $\cdots$ & $\cdots$ & $\cdots$ & $12.73\pm0.16$ & 0.61  & 13 \\
NGC6441  & V6   & 21.365	& 16.117   & 14.885   & 13.231   & 12.16    & 11.64    & 11.49    & $12.73\pm0.16$ & 0.61  & 13 \\
M5       & V42  & 25.735	& 11.82    & 11.659   & 10.740   & 10.16    & 9.85     & 9.82     & $7.48\pm0.06$ & 0.04  & 1, 14 \\
M5       & V84  & 26.87	        & 12.11    & 11.287   & 10.451   & 10.20    & 9.80     & 9.71     & $7.48\pm0.06$ & 0.04  & 1, 14 \\
NGC6453  & V2   & 27.1954	& $\cdots$ & 14.231   & 12.375   & 11.35    & 10.75    & 10.59    & $10.07\pm0.22$ & 0.66  &  17 \\
NGC5139  & V1   & 29.3479	& 11.488   & 10.829   & 10.058   & 9.40     & 9.05     & 8.99     & $5.43\pm0.05$ & 0.13  & 3 \\
NGC6453  & V1   & 31.0476	& $\cdots$ & 14.601   & 12.789   & 11.51    & 10.85    & 10.66    & $10.07\pm0.22$ & 0.66  & 17 \\
M2       & V11  & 33.4	        & 12.67    & 12.11    & $\cdots$ & 10.87    & 10.53    & 10.44    & $11.69\pm0.11$ & 0.04  & 9 \\
NGC5986  & V13  & 40.62	        & $\cdots$ & $\cdots$ & $\cdots$ & 10.90    & 10.22    & 10.07    & $10.54\pm0.13$ & 0.34  &  \\
M56      & V6   & 45.0	        & 13.7     & 12.9     & $\cdots$ & 10.86    & 10.37    & 10.21    & $10.43\pm0.14$ & 0.25  & 18 \\
M28      & V17  & 48.0	        & $\cdots$ & $\cdots$ & $\cdots$ & 9.55     & 8.95     & 8.75     & $5.37\pm0.10$ & 0.49  &  \\
NGC6569  & V16  & 87.5	        & 16.55    & $\cdots$ & $\cdots$ & 10.56    & 9.74     & 9.45     & $10.53\pm0.26$ & 0.43  & 10  \\
  \enddata
  \tablenotetext{a}{Distance of the globular clusters adopted from \citet{baumgardt2021}.}
  \tablenotetext{b}{Reddening returned from the ``SFD'' dust map \citep{sfd1998}.}
  \tablenotetext{c}{Sources for the $BVI$-band mean magnitudes: 1 = \citet{af2016}; 2 = \citet{af2020}; 3 = \citet{braga2020}; 4 = \citet{clement1978}; 5 = \citet{clement1980}; 6 = \citet{conteras2010}; 7 = \citet{cp2018}; 8 = \citet{corwin2008}; 9 = \citet{demers1969}; 10 = \citet{hl1985}; 11 = \citet{kopacki2013}; 12 = \citet{kunder2013}; 13 = \citet{pritzl2003}; 14 = \citet{rabidoux2010}; 15 = \citet{rozyczka2018}; 16 = \citet{skottfelt2015}; 17 = \citet{udalski2018}; 18 = \citet{wehlau1985}; 19 = \citet{wehlau1990a}; 20 = \citet{wehlau1990}.}
\end{deluxetable*}

\subsection{Updated the PL Relations} \label{sec4.2}

\begin{figure*}
  \gridline{
    \fig{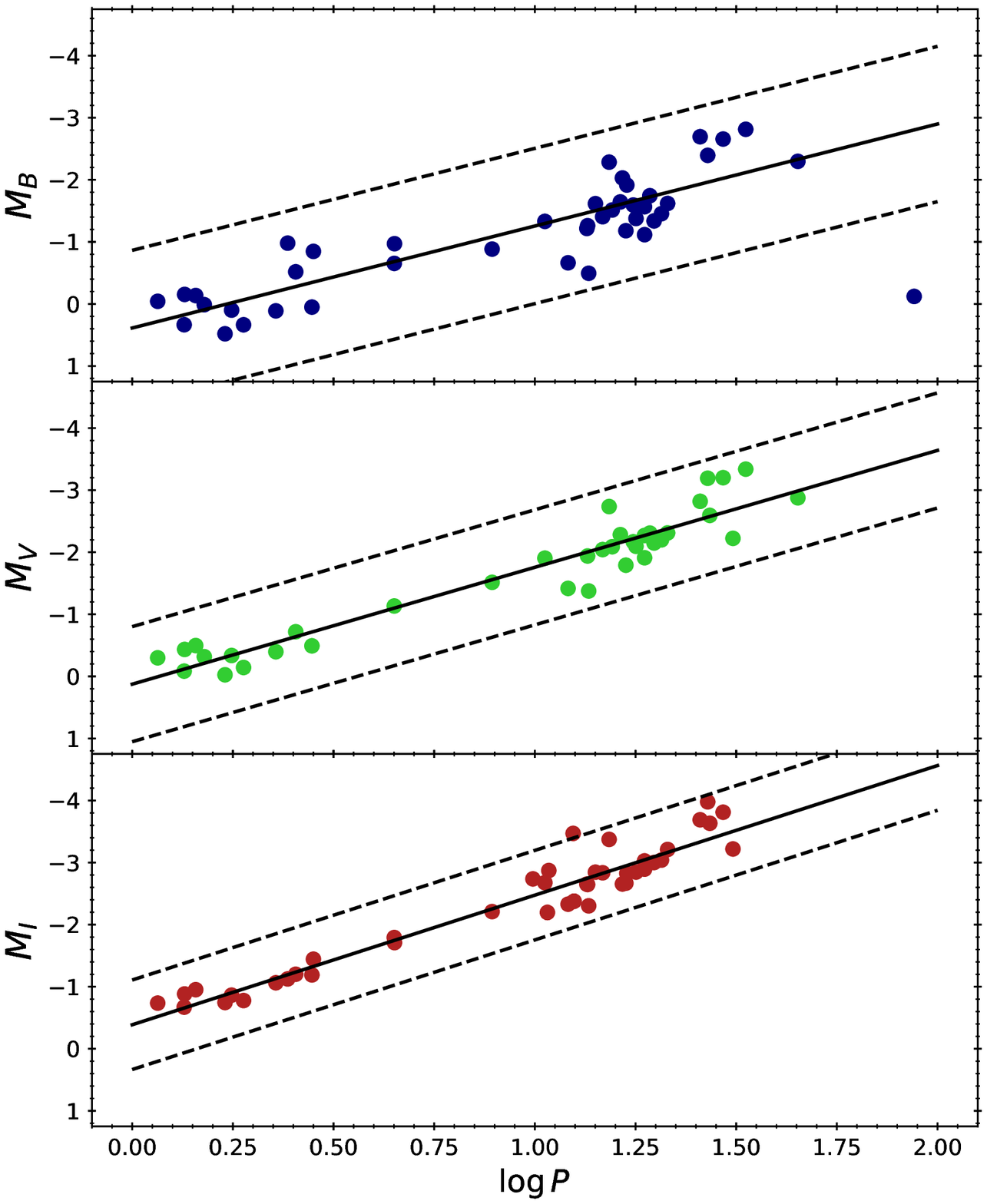}{0.32\textwidth}{$BVI$-band PL relations}
    \fig{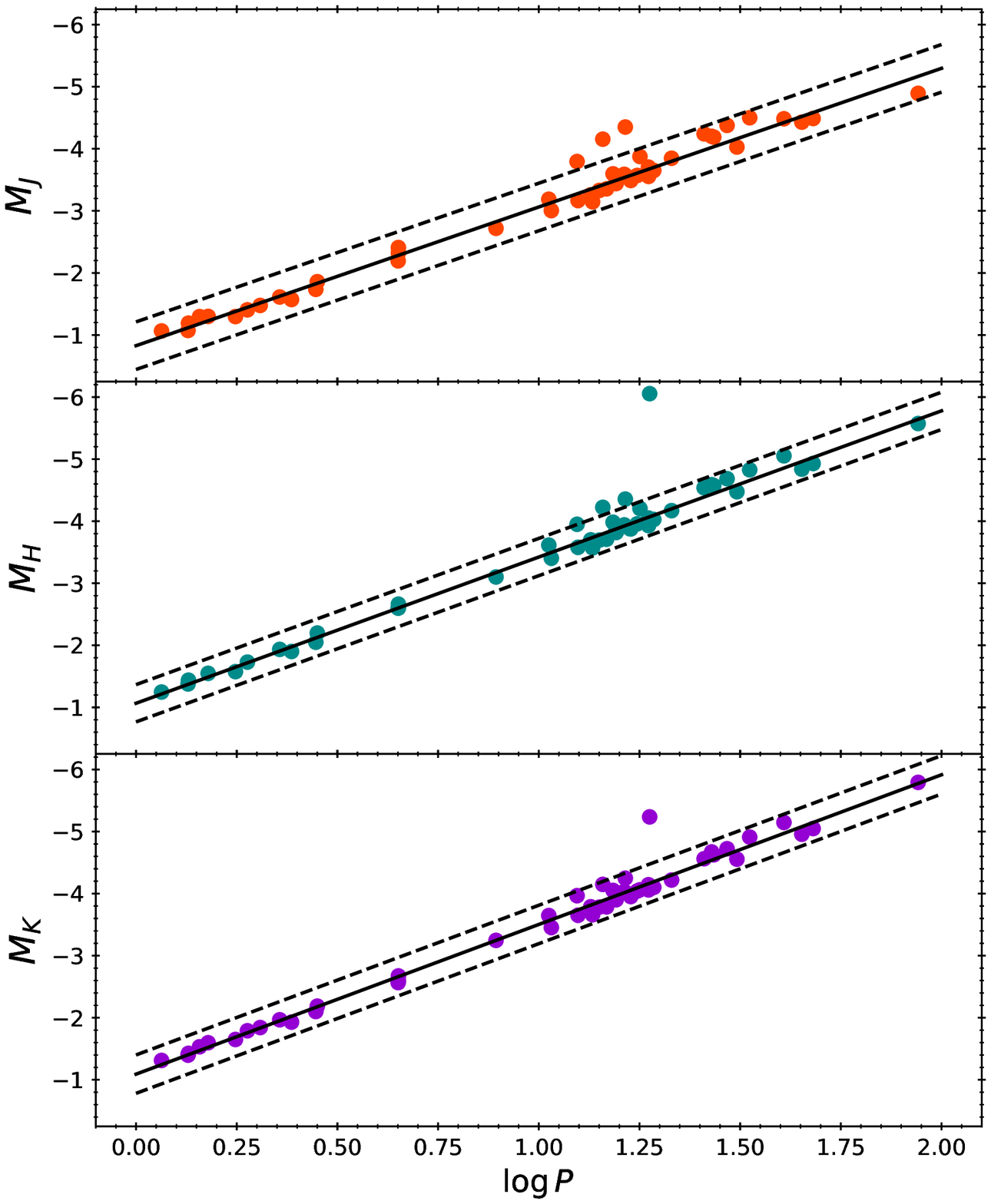}{0.32\textwidth}{$JHK$-band PL relations}
    \fig{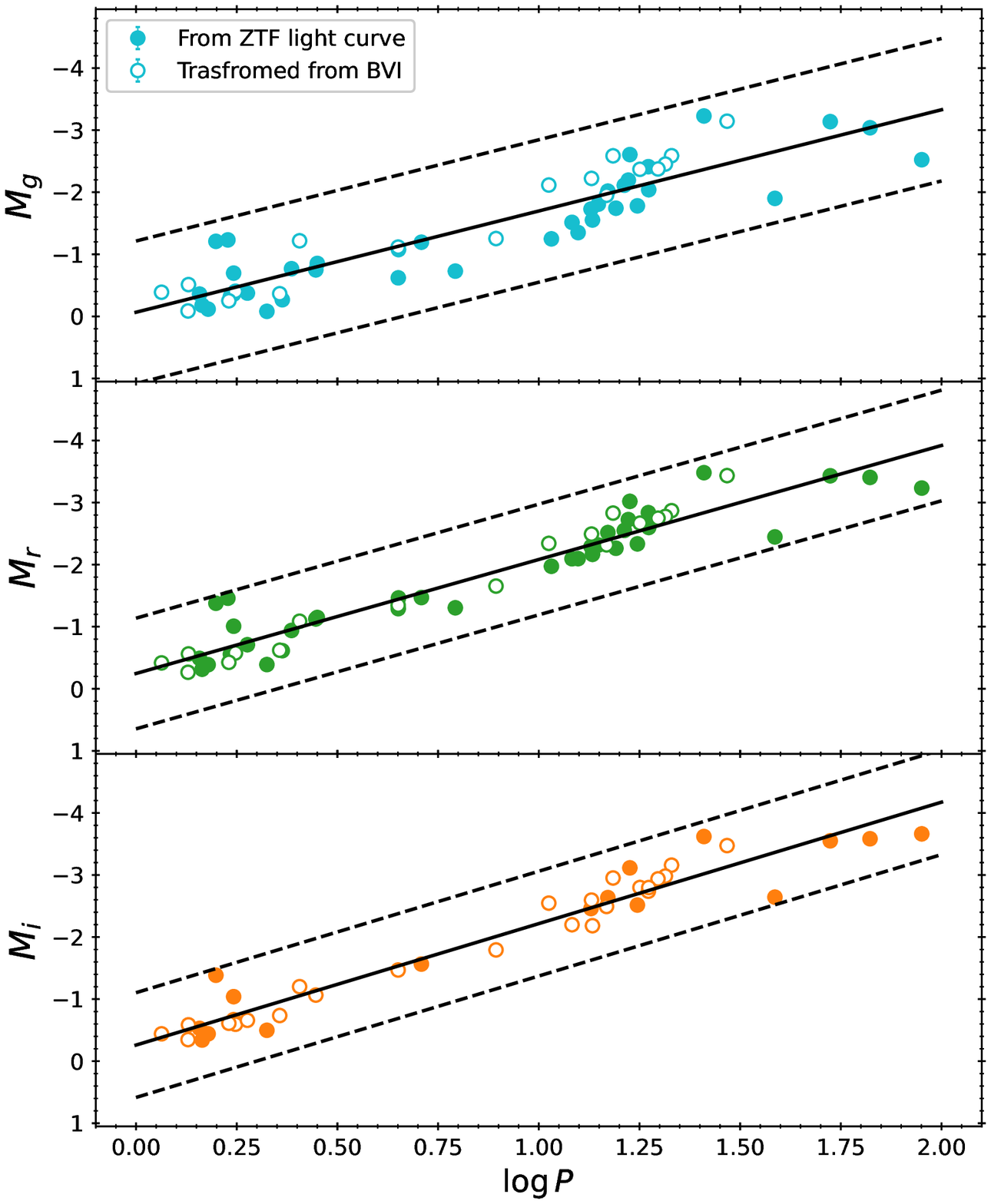}{0.32\textwidth}{$gri$-band PL relations}
  }
  \caption{The updated multi-band PL relations, where the best-fit PL relations are shown in solid lines (see Table \ref{tab_pl}), and the dashed lines represent the $\pm3\sigma$ of the best-fit PL relations (hence, data points outside the $\pm 3\sigma$ range are rejected in the fitting). The $BVIJHK$-band PL relations were updated from \citet{bhardwaj2022} by using the homogeneous distance from \citet{baumgardt2021}. The updated $gri$-band PL relations were derived from combining two samples of TIIC: those with ZTF data as listed in Table \ref{tab_t2cep}, and the TIIC from B22 sample that are not included in Table \ref{tab_t2cep} (after transformed to the $gri$ bands). See Section \ref{sec4.2} for more details. Note that the error bars are smaller than the size of the symbols.}
  \label{fig_pl}
\end{figure*}

\begin{deluxetable}{ccrcc}
  \tabletypesize{\scriptsize}
  \tablecaption{The Derived Period-Luminosity Relations for TIIC in the Globular Clusters \label{tab_pl}}
  \tablewidth{0pt}
  \tablehead{
    \colhead{Band} &
    \colhead{$a$} &
    \colhead{$b$} &
    \colhead{$\sigma$} &
    \colhead{$N$} 
  }
  \startdata  
  $B$  & $-1.64\pm0.14$ &  $0.39\pm0.14$ & 0.42 & 42 \\
  $V$  & $-1.88\pm0.10$ &  $0.13\pm0.11$ & 0.31 & 37 \\
  $I$  & $-2.09\pm0.08$ & $-0.39\pm0.08$ & 0.24 & 41 \\
  \hline
  $J$  & $-2.23\pm0.04$ & $-0.83\pm0.04$ & 0.13 & 45 \\
  $H$  & $-2.36\pm0.03$ & $-1.07\pm0.04$ & 0.10 & 43 \\
  $K$  & $-2.41\pm0.03$ & $-1.09\pm0.03$ & 0.10 & 48 \\
  \hline
  $g$  & $-1.63\pm0.10$ & $-0.07\pm0.10$ & 0.38 & 55 \\
  $r$  & $-1.84\pm0.08$ & $-0.25\pm0.08$ & 0.30 & 55 \\
  $i$  & $-1.96\pm0.08$ & $-0.26\pm0.08$ & 0.28 & 41 \\  
  \enddata
  \tablecomments{The PL relation takes the form of $m=a\log P + b$, and $\sigma$ is the dispersion of the fitted PL relation. $N$ represents the number of TIIC used in the fitting.}
\end{deluxetable}

We updated the $BVIJHK$-band PL relations for the TIIC sample compiled in \citet[hereafter B22 sample]{bhardwaj2022} by adopting the homogeneous distance from \citet{baumgardt2021} to their host globular clusters. We have also adopted the homogeneous reddening $E(B-V)$ queried from the same all-sky ``SFD'' dust map \citep{sfd1998}, using the {\tt dustmaps} code, to the TIIC in B22 sample. The compiled $BVIJHK$-band mean magnitudes (whenever available), as well as the adopted distances and reddenings, for the B22 sample are presented in Table \ref{tab_b22}. Mean magnitudes in the $BVI$-band were adopted from various sources as listed in the last column of Table \ref{tab_b22}. For $JHK$-band mean magnitudes, majority of them were taken from \citet{matsunaga2006} except for V34 in M15 \citep{bhardwaj2021} and V43, V60, V61, and V92 in NGC 5139 \citep{braga2020}. We excluded V1 in M10 and V8 in M79 from the B22 sample for the reasons mentioned in Section \ref{sec2.1}.

The $JHK$ photometry from the aforementioned three studies was homogeneously calibrated to 2MASS \citep[2 Micron All Sky Survey,][]{2mass2006} system. However, the optical photometric data are very heterogeneous and were taken from several different studies as evident from the last column of Table \ref{tab_b22}. Since most of the mean magnitudes do no have their associated photometric measurement errors and are likely to suffer from systematic uncertainties, we adopt an error of 0.05 magnitudes on the mean magnitudes. The available mean magnitudes listed in Table \ref{tab_b22} were converted to absolute magnitudes using the adopted distances. Extinction corrections on $BVIJHK$-band mean magnitudes were done using $A_{BVIJHK}=R_{BVIJHK}E(B-V)$, where $R_{BVIJHK}=\{3.626,\ 2.742,\ 1.505,\ 0.793,\ 0.469,\ 0.303\}$ \citep{schlafly2011,green2019}. We then fit the PL relations using an iterative $3\sigma$-clipping linear regression (where $\sigma$ is the dispersion of the regression), implemented in {\tt astropy}, to exclude a few obvious outliers. The updated $BVIJHK$-band PL relations are shown in Figure \ref{fig_pl} and provided in Table \ref{tab_pl}. 

There are 33 TIIC in the B22 sample that are not included in Table \ref{tab_t2cep}. Majority of these TIIC were located at south of $\delta_{J2000} = -30^\circ$ (i.e. outside the ZTF footprint), and the remaining TIIC either did not have ZTF light curve data or were excluded (e.g. due to blending). The $BVI$-band mean magnitudes for these TIIC, whenever available, were transformed to the $gri$-band using the transformations provided in \citet{tonry2012}. Extinction corrections were done using the {\tt Bayerstar2019} 3D reddening map if available, else the ``SFD'' dust map was used together with the conversion of $E=E(B-V)/0.884$ (see footnote \ref{fn2}). Similarly, there are 25 common TIIC in B22 sample and Table \ref{tab_t2cep},\footnote{We checked the consistency of transformed $gri$-band mean magnitudes using the 25 common TIIC in B22 sample and Table \ref{tab_t2cep}. The averaged differences of $m_{ZTF}-m_T$ in the $gri$-band are $0.014$, $-0.123$, and $-0.020$~mag, respectively, where $m_{ZTF}$ and $m_T$ represent the ZTF and the transformed mean magnitudes. The corresponding standard deviations in the $gri$-band are $0.092$, $0.163$, and $0.160$~mag, respectively. Note that after removing an extreme outlier, the number of TIIC in both samples with mean magnitudes to calculate the averaged difference is 16 for the $gr$-band, and 3 for the $i$-band. The revised $r$-band PL relation, $M_r=-1.83(\pm0.08)\log P - 0.29(\pm0.08)$ with $\sigma=0.31$~mag, is consistent with Table \ref{tab_pl} after taking the averaged differences of $-0.12$~mag into account.} the $BVI$-band mean magnitudes from B22 sample were transformed to the $i$-band for those TIIC without the $i$-band data. Open circles in the right panels of Figure \ref{fig_pl} represent the TIIC in B22 sample transformed from the $BVI$-band photometry.

Combining the TIIC in Table \ref{tab_t2cep} and those transformed from B22 sample, we derived the updated $gri$-band PL relation, using the same iterative $3\sigma$-clipping linear regression. The results are listed in the bottom part of Table \ref{tab_pl}. With the updated PL relations, derived using the homogeneous distances, consistent PL relations were found between the $BVI$-band PL relations and the $gri$-band PL relations, as demonstrated in the right-panel of Figure \ref{fig_compare}.

\begin{figure}
  \epsscale{1.1}
  \plotone{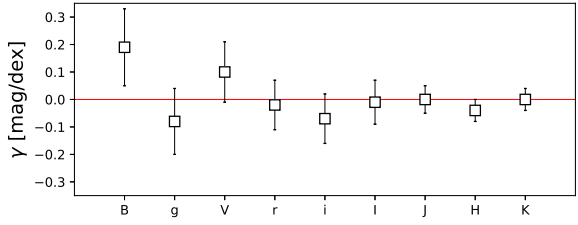}
  \caption{Slopes of the fitted linear regressions, $\gamma$, to the plots for PL residuals vs. $\mathrm{[Fe/H]}$ as a function of filters. The red line indicates the case of $\gamma=0$.}
  \label{fig_feh}
\end{figure}

Most of the previous studies have suggested that the PL relations for TIIC are insensitive to metallicity \citep[for examples, see][and reference therein]{matsunaga2006,dic2007,matsunaga2009,ciechanowska2010,ripepi2015,groenewegen2017,braga2018,bhardwaj2020,bhardwaj2022}. In contrast, significant metallicity terms were found for the $UB$-band and $JHK$-band PL relations from theoretical work of \citet{das2021} and empirical investigations of \citet{wiegorski2021}, respectively. Following \citet{matsunaga2006} and \citet{wiegorski2021}, we fit a linear regression to the residuals of PL relations as a function of metallicity for our sample of TIIC, where the metallicities, $\mathrm{[Fe/H]}$ for the host globular clusters, were taken from the GOTHAM (GlObular clusTer Homogeneous Abundances Measurements) survey\footnote{\url{http://www.sc.eso.org/~bdias/files/dias+16\_MWGC.txt}} \citep{dias2015, dias2016a, dias2016b, vasquez2018}. Metallicity of these host globular clusters ranged from $-2.27$~dex (M15) to $-0.47$~dex (NGC6441). Slopes of these linear regressions, denoted as $\gamma$, as a function of filters are displayed in Figure \ref{fig_feh}. Except in $B$-band, the values of $\gamma$ are consistent with zero in all other filters, implying the corresponding PL relations are insensitive to metallicity. This is consistent with the theoretical predictions of \citet{das2021}. For $B$-band, fitting a period-luminosity-metallicity relation to the data yields:

\begin{eqnarray}
  M_B & = & 0.68 (\pm0.25) - 1.67 (\pm0.14)\log P  \nonumber \\
 &  & + 0.19 (\pm0.14) \mathrm{[Fe/H]},\ \  \sigma=0.41.\nonumber
\end{eqnarray}

\section{The Multi-Band Relations} \label{sec5}

\begin{figure*}
  \epsscale{1.1}
  \gridline{
    \fig{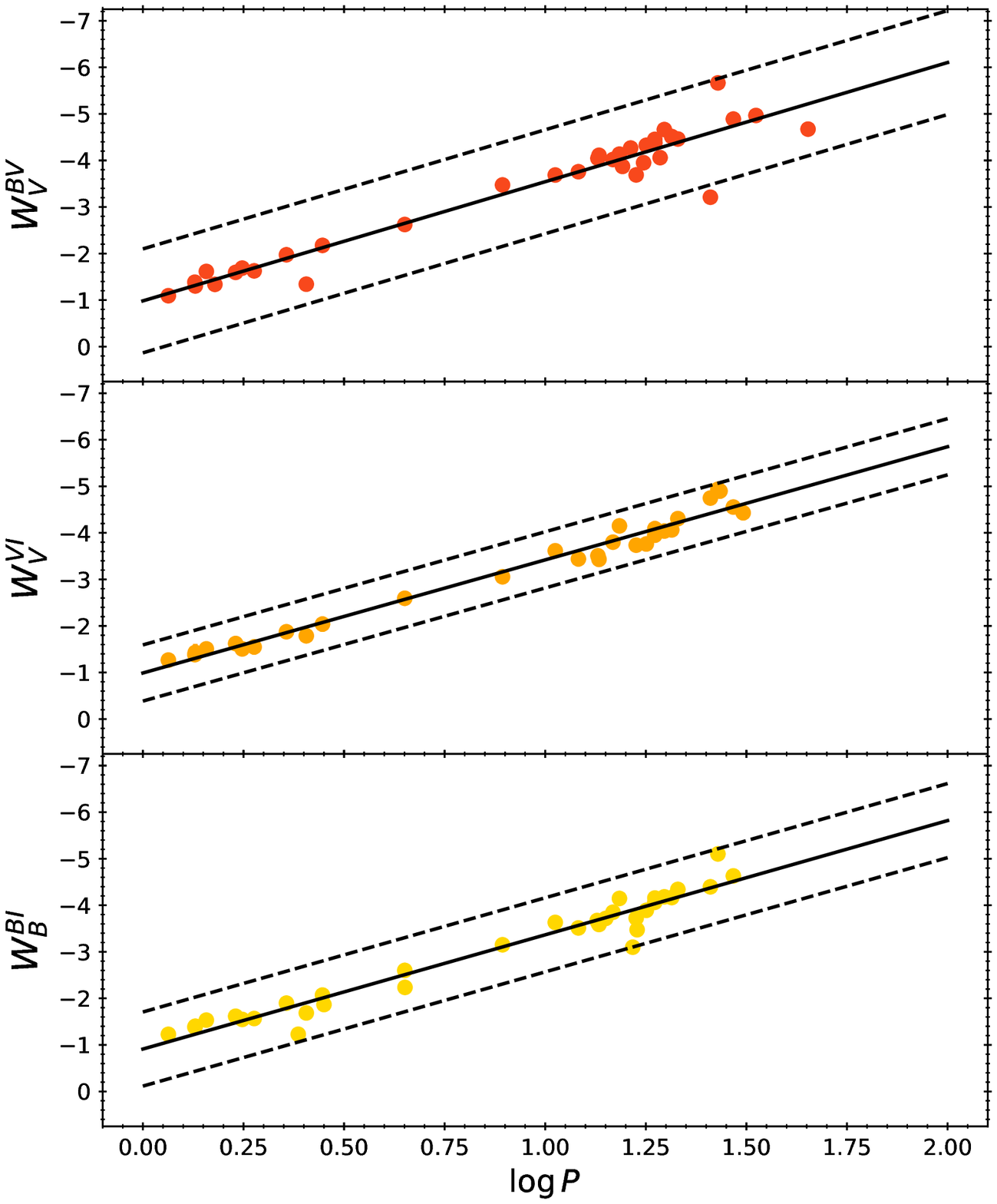}{0.32\textwidth}{For $BVI$-band}
    \fig{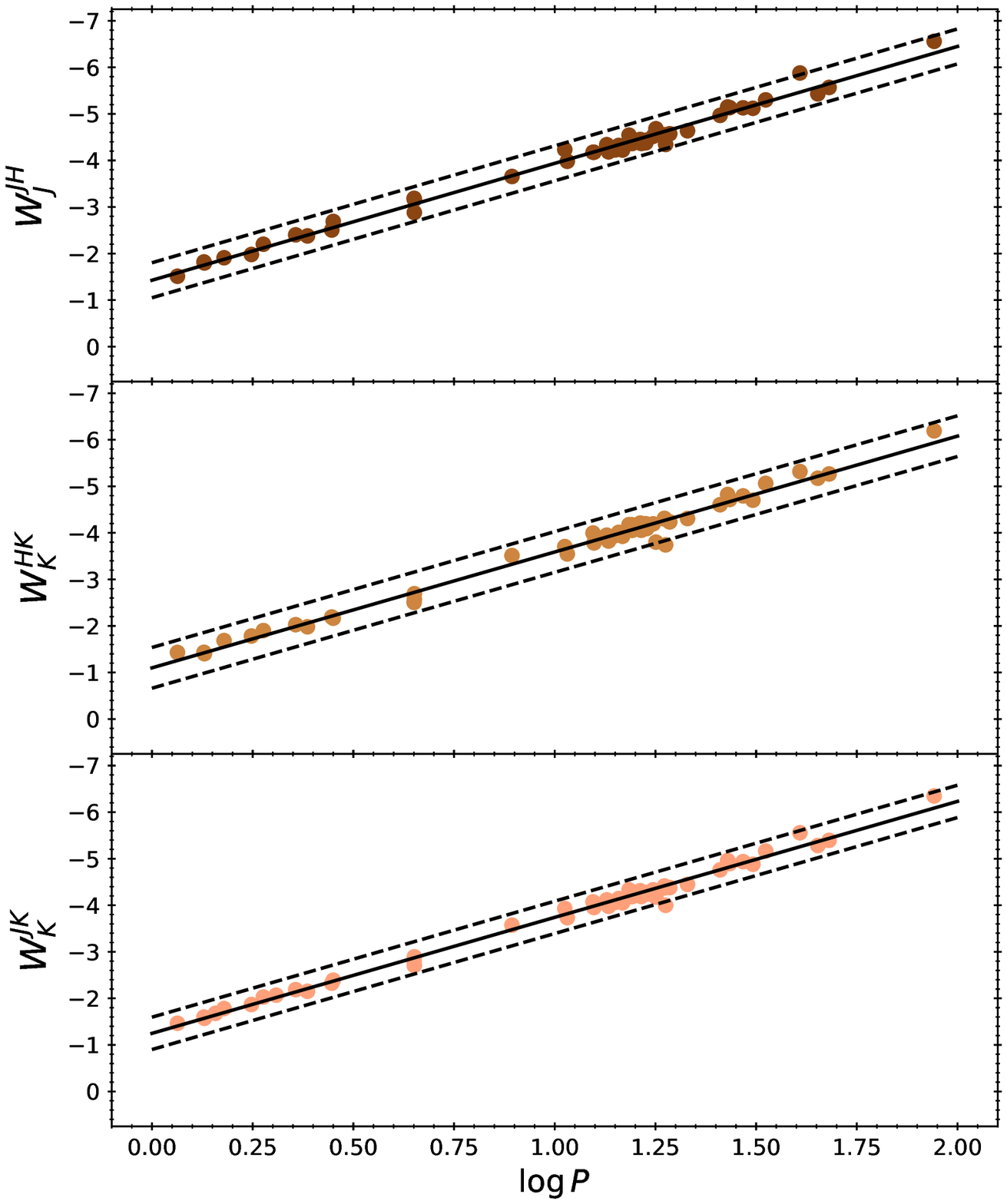}{0.32\textwidth}{For $JHK$-band}
    \fig{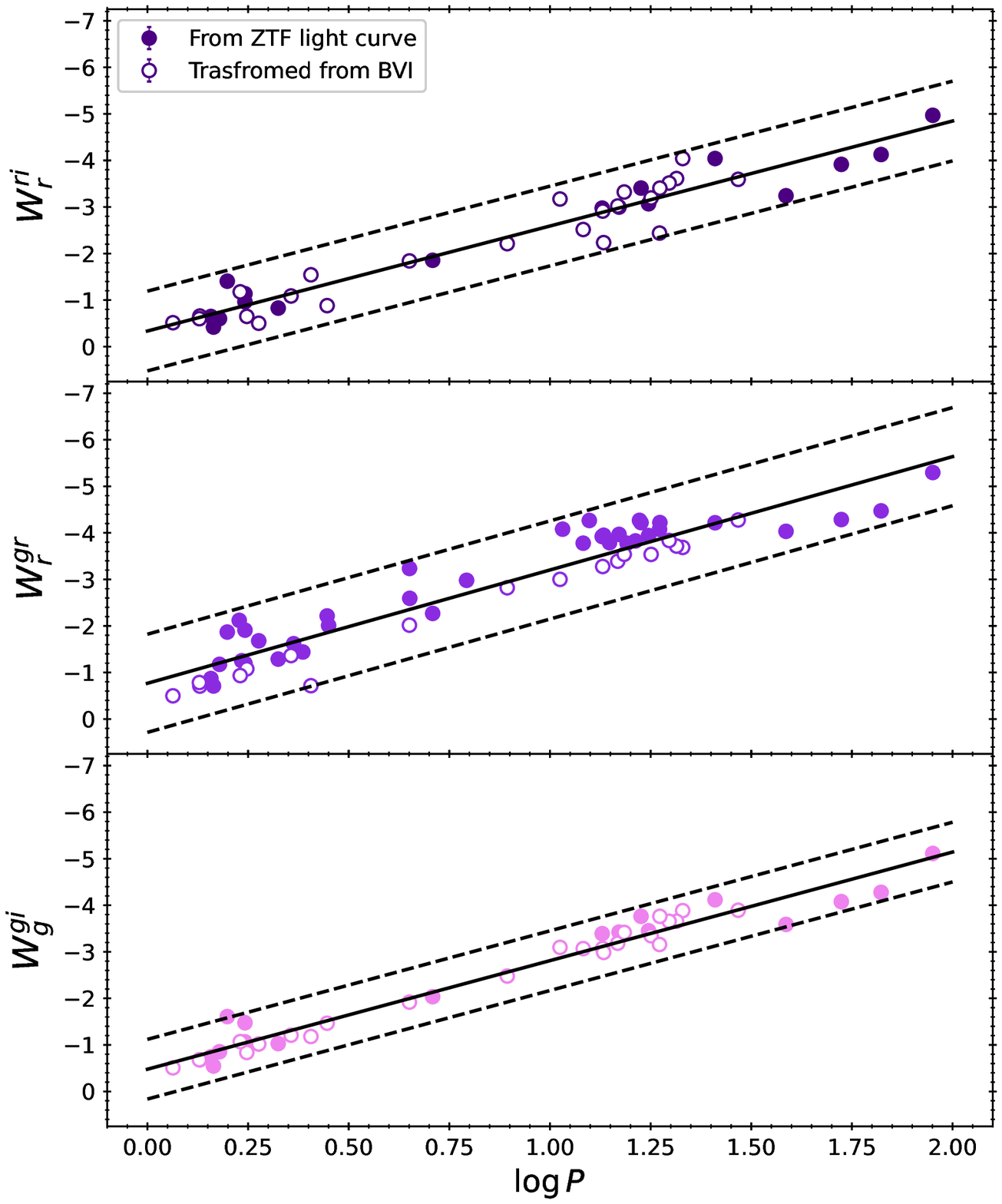}{0.32\textwidth}{For $gri$-band}
  }
  \caption{Same as Figure \ref{fig_pl}, but for the PW relations.}
  \label{fig_pw}
\end{figure*}

\begin{deluxetable}{lcrcc}
  \tabletypesize{\scriptsize}
  \tablecaption{The Derived Period-Wesenheit Relations for TIIC in the Globular Clusters \label{tab_pw}}
  \tablewidth{0pt}
  \tablehead{
    \colhead{Wesenheit Index} &
    \colhead{$a$} &
    \colhead{$b$} &
    \colhead{$\sigma$} &
    \colhead{$N$} 
  }
  \startdata  
  $W^{BV}_V = V - 3.102 (B-V)$ & $-2.62\pm0.05$ & $-1.00\pm0.05$ & 0.13 & 30 \\
  $W^{VI}_V = V - 2.217 (V-I)$ & $-2.43\pm0.07$ & $-0.99\pm0.08$ & 0.20 & 30 \\
  $W^{BI}_B = B - 1.710 (B-I)$ & $-2.42\pm0.05$ & $-0.98\pm0.05$ & 0.14 & 31 \\ 
  \hline
  $W^{JH}_J = J - 2.448 (J-H)$ & $-2.49\pm0.03$ & $-1.44\pm0.04$ & 0.11 & 46 \\
  $W^{HK}_K = K - 1.825 (H-K)$ & $-2.51\pm0.03$ & $-1.10\pm0.04$ & 0.11 & 45 \\
  $W^{JK}_K = K - 0.618 (J-K)$ & $-2.46\pm0.03$ & $-1.28\pm0.03$ & 0.08 & 46 \\ 
  \hline
  $W^{ri}_r = r - 4.051 (r-i)$ & $-2.26\pm0.10$ & $-0.34\pm0.10$ & 0.34 & 41 \\
  $W^{gr}_r = r - 2.905 (g-r)$ & $-2.43\pm0.11$ & $-0.77\pm0.11$ & 0.42 & 55 \\
  $W^{gi}_g = g - 2.274 (g-i)$ & $-2.33\pm0.07$ & $-0.48\pm0.07$ & 0.26 & 41 \\
  \enddata
  \tablecomments{The PW relation takes the form of $W=a\log P + b$, and $\sigma$ is the dispersion of the fitted relation. $N$ represents the number of TIIC used in the fitting.}
\end{deluxetable}

\begin{deluxetable}{ccrcc}
  \tabletypesize{\scriptsize}
  \tablecaption{The Derived Period-Color and Period-Q-index Relations for TIIC in the Globular Clusters \label{tab_pcq}}
  \tablewidth{0pt}
  \tablehead{
    \colhead{Color} &
    \colhead{$a$} &
    \colhead{$b$} &
    \colhead{$\sigma$} &
    \colhead{$N$} 
  }
  \startdata  
  $(B-V)$  & $0.24\pm0.04$ & $0.35\pm0.04$ & 0.11 & 34 \\
  $(V-I)$  & $0.26\pm0.04$ & $0.50\pm0.04$ & 0.11 & 30 \\
  $(B-I)$  & $0.50\pm0.11$ & $0.75\pm0.11$ & 0.31 & 35 \\  
  \hline
  $(J-H)$  & $0.08\pm0.02$ & $0.27\pm0.02$ & 0.05 & 41 \\
  $(H-K)$  & $0.05\pm0.01$ & $0.02\pm0.01$ & 0.04 & 42 \\
  $(J-K)$  & $0.14\pm0.02$ & $0.29\pm0.02$ & 0.06 & 42 \\  
  \hline
  $(g-r)$  & $0.21\pm0.04$ & $0.18\pm0.04$ & 0.15 & 55 \\
  $(r-i)$  & $0.09\pm0.02$ & $0.04\pm0.02$ & 0.06 & 39 \\
  $(g-i)$  & $0.29\pm0.04$ & $0.17\pm0.04$ & 0.15 & 41 \\
  \hline
  $Q_{BVI}$ & $0.12\pm0.03$ &$-0.04\pm0.03$ & 0.07 & 27 \\
  $Q_{JHK}$ & $0.02\pm0.02$ & $0.21\pm0.03$ & 0.08 & 45 \\  
  $Q_{gri}$ & $0.06\pm0.05$ & $0.11\pm0.05$ & 0.17 & 41 \\  
  \enddata
  \tablecomments{The PC and PQ relations take the form of $c=a\log P + b$ (where $c$ is for colors or $Q$-index), and $\sigma$ is the dispersion of the fitted relation. $N$ represents the number of TIIC used in the fitting.}
\end{deluxetable}

\begin{figure*}
  \epsscale{1.1}
  \gridline{
    \fig{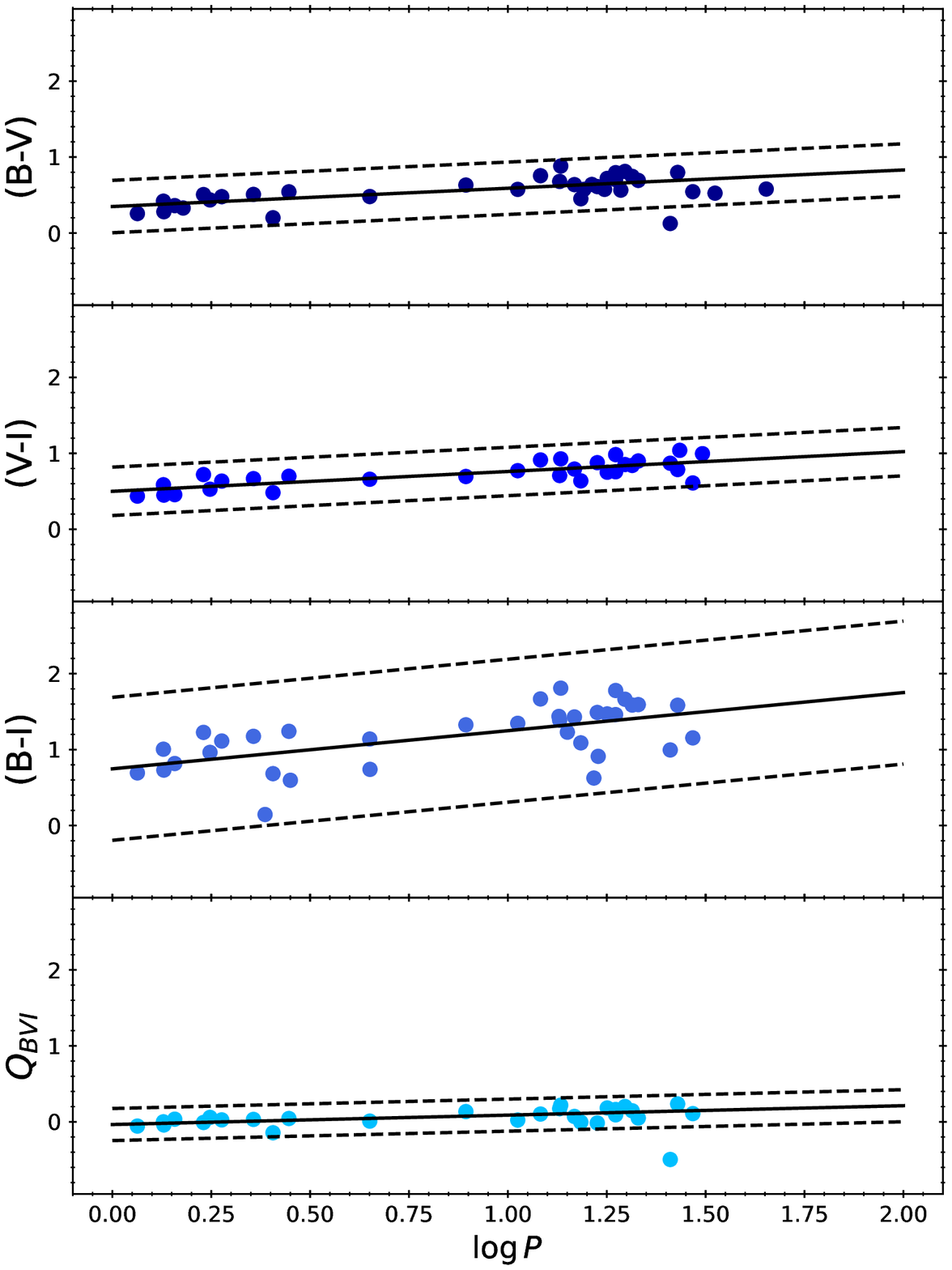}{0.32\textwidth}{For $BVI$-band}
    \fig{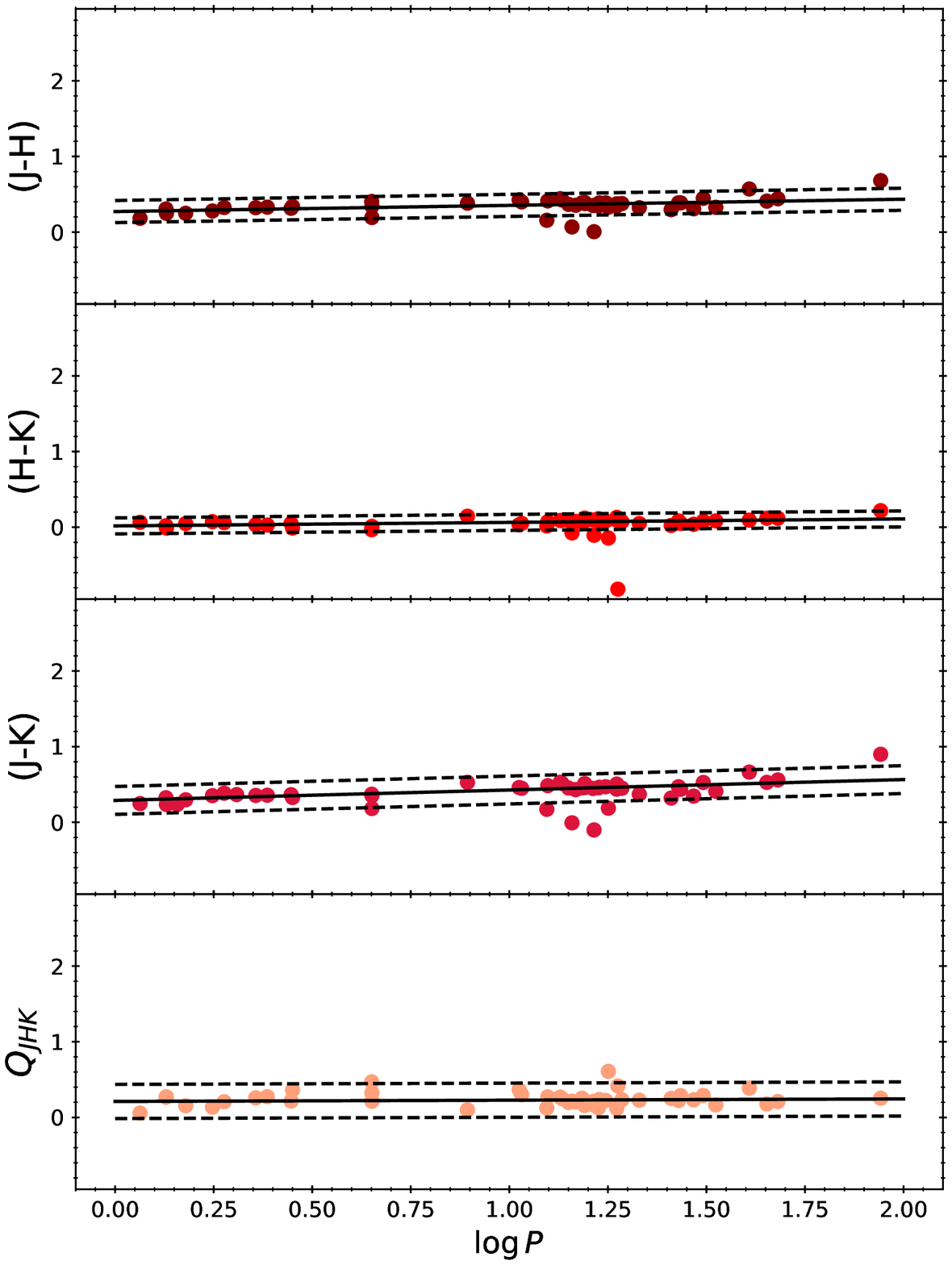}{0.32\textwidth}{For $JHK$-band}
    \fig{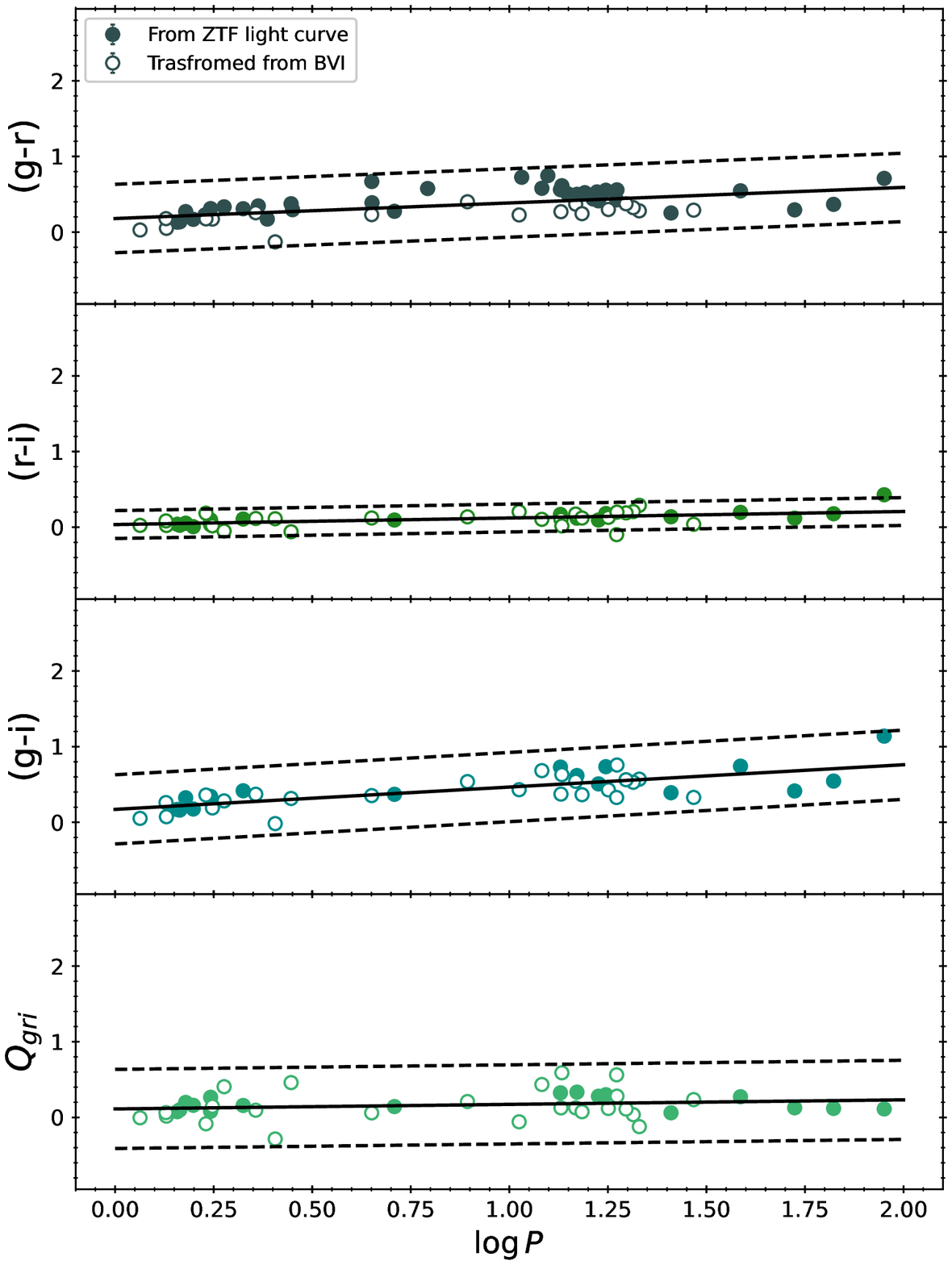}{0.32\textwidth}{For $gri$-band}
  }
  \caption{Same as Figure \ref{fig_pl}, but for the reddening-corrected PC relations (top three panels) and the reddening-free PQ relation (bottom panel). Scales on the $y$-axis were intended to be the same in all panels, such that the PC/PQ relations can be compared.}
  \label{fig_pcq}
\end{figure*}

In addition to PL relations, the updated B22 samples can be used to derive the period-Wesenheit (PW), period-color (PC), and the period-$Q$-index (PQ) relations in the $BVIJHK$-band. The Wesenheit index, $W$, is analog to magnitude but it is extinction-free by construction \citep{madore1982,madore1991}. Similarly, the $Q$-index is analog to color but reddening-free by construction, inspired from the classical work of \citet[][who defined the $Q$-index in $UBV$-band]{johnson1953}. The combined sample of TIIC listed in Table \ref{tab_t2cep} and those photometrically transformed from the B22 sample can also be used to derive the $gri$-band PW, PC, and PQ relations. The $gri$-band Wesenheit indices were defined in \citet{ngeow2021}, while the various $BVIJHK$-band Wesenheit indices are defined in Table \ref{tab_pw}. For the PQ relations, we have $Q_{BVI} = (B-V) - 0.715(V-I)$ and $Q_{JHK} = (J-H) - 1.952(H-K)$ in the $BVIJHK$-band, while the $gri$-band $Q$-index was adopted from \citet{ngeow2022} as $Q_{gri} = (g-r) - 1.395(r-i)$. The fitted PW and PC/PQ relations are summarized in Table \ref{tab_pw} and \ref{tab_pcq}, respectively, as well as presented in Figure \ref{fig_pw} and \ref{fig_pcq}.

The $(H-K)$ and $(r-i)$ PC relations have relatively flat PC slopes with zero-points almost consistent with zero. These explain why the pairs of $HK$-band and $ri$-band PL relations are quite similar, especially their PL zero-points are identical within the uncertainties (see Table \ref{tab_pl}). We also see that the redder colors, in $JHK$-band and in $(r-i)$ color, tend to have the smaller PC dispersion. In contrast, the $(B-I)$ PC relation displays the largest dispersion among all the PC relations. In case of the PQ relations, slopes for both of the $Q_{JHK}$ and $Q_{gri}$ PQ relations are statistically consistent with zero, in contrast to the RR Lyrae \citep{ngeow2022}. The $Q_{BVI}$ PQ relation is also much shallower than the $BVI$-band PC relations, and has the smallest dispersion among the three PQ relations.

\section{Comparison with M31 TIIC} \label{sec6}

\begin{figure*}
  \epsscale{1.1}
  \plottwo{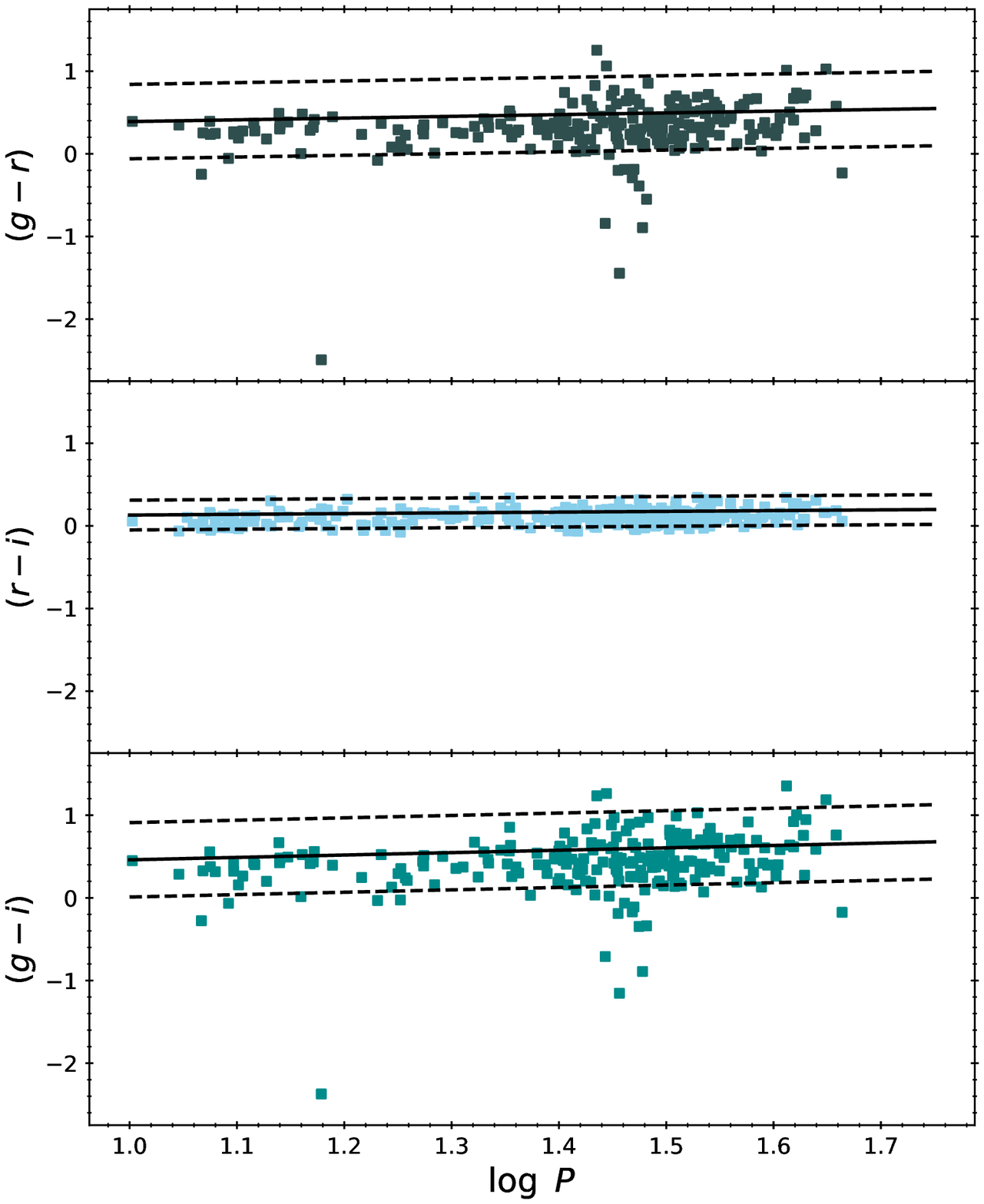}{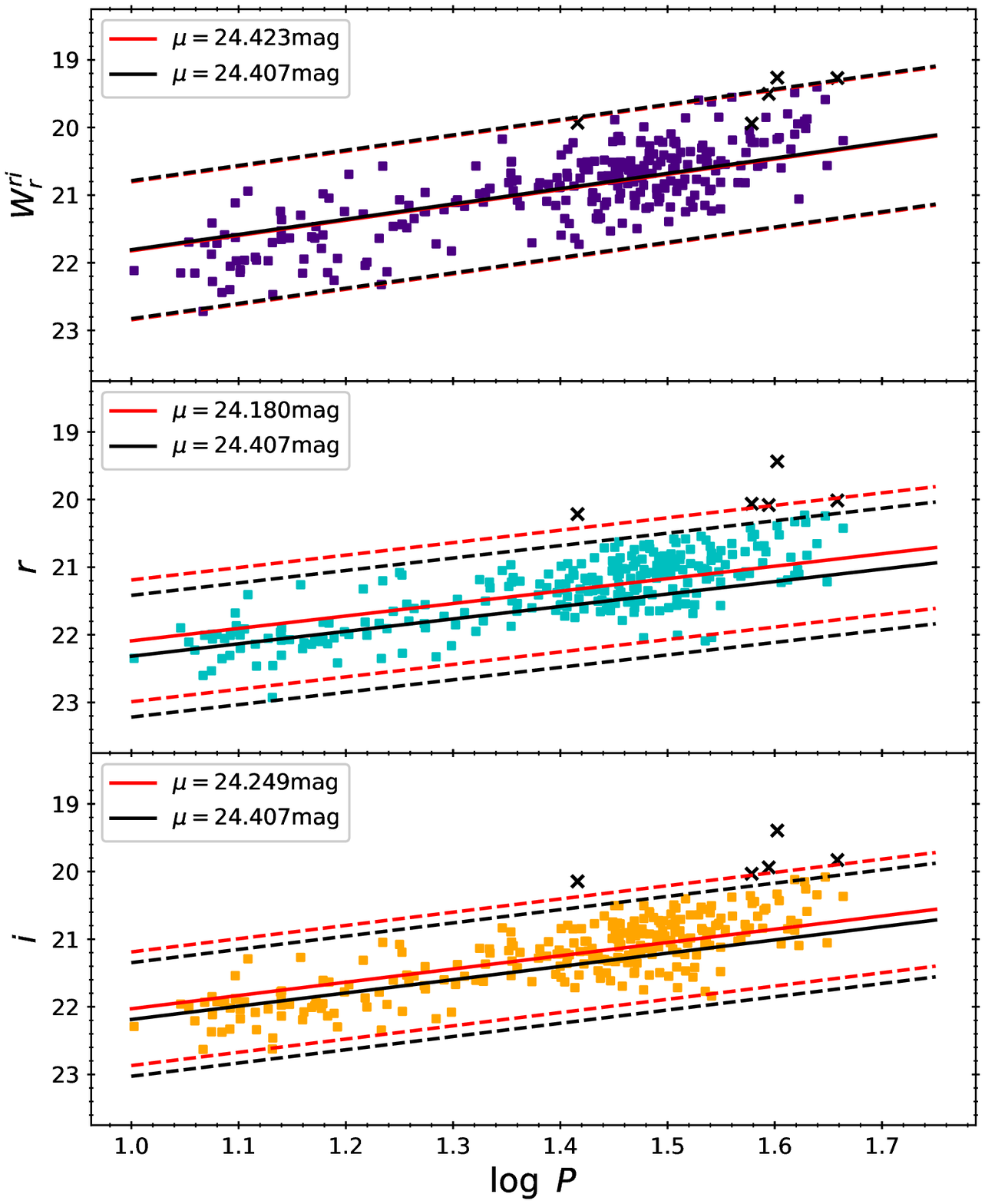}
  \caption{{\bf Left Panel:} PC relations for the M31 TIIC, where the colors of the TIIC have been reddening-corrected \citep{kodric2018}. The solid lines are the PC relations given in Table \ref{tab_pcq}, together with the $\pm3\sigma$ boundaries shown as dashed lines. {\bf Right Panel:} The PW relation (top-right panel) for the $W^{ri}_r$ Wesenheit-index, and the extinction-corrected $ri$-band PL relations (middle-right and bottom-right panels) for the M31 TIIC. Crosses represent the rejected TIIC as described in the text (see Section \ref{sec6}). Similar to the left panels, the solid lines are the PL/PW relations given in Table \ref{tab_pl} and \ref{tab_pw}, respectively, shifted vertically with the distance modulus ($\mu$) of M31, and the dashed lines are the corresponding $\pm3\sigma$ boundaries. Black lines are the shifted PL/PW relations by adopting the same $\mu=24.407$~mag \citep{li2021}. The red lines represent the PL/PW relations after shifting the $\mu$ determined from fitting the data to the PL/PW relations given in Table \ref{tab_pl} and \ref{tab_pw}. In both panels, error bars are omitted for clarity.}
  \label{fig_m31pc}
\end{figure*}

The Pan-STARRS1 survey of Andromeda, known as the PAndromeda project, reported a finding of 278 TIIC in the (halo of) M31 galaxy \citep{kodric2018}. This sample of M31 TIIC can be used to test the applicability of our derived PL/PW relations. Numerous distance measurements to M31, via various techniques and distance indicators, can be found in the literature. \citet{dgb2014} summarized the distance estimates prior to 2013 and recommended a distance modulus of $\mu=24.46\pm0.10$~mag to M31. A latest distance measurement to M31 can be found in \citet{li2021}, who give $\mu=24.407\pm0.032$~mag based on the Hubble Space Telescope observations of classical Cepheids. 

\citet{kodric2018} provided the pulsation periods as well as the extinction-corrected $gri$-band mean magnitudes for these sample of M31 TIIC. We first removed six TIIC that have errors on the periods which are larger than 1~day (or fractional error larger than 1\%; the rest of the TIIC have fractional errors that are less than 0.64\% in period). The reddening-corrected colors for the remaining 272 TIIC were plotted against their logarithmic period in the left panel of Figure \ref{fig_m31pc}, overlaid with the PC relations taken from Table \ref{tab_pcq}. The $(r-i)$ colors for the M31 TIIC are remarkable in good agreement with the $(r-i)$ PC relation derived from our sample of TIIC located in the globular clusters. In contrast, outliers can be seen on the $(g-r)$ and $(g-i)$ PC relations, suggesting there could be some problems in the $g$-band. Indeed, the $g$-band observations were $\sim5$ to $\sim10$ times less than the $ri$-band \citep{kodric2018}, such that the $g$-band light curves do not have quality as good as in other two bands. As a result, out of the remaining 272 TIIC, 50 of them do not have mean $g$-band magnitudes, and 161 of them carry a non-zero bit flag \citep[see Table 2 of][]{kodric2018} indicating there are some problems associated with the $g$-band data. For these reasons, we only focused on the $ri$-band mean magnitudes for this sample of TIIC in the subsequent analysis.

Right panels of Figure \ref{fig_m31pc} present the $ri$-band PL/PW relations for the M31 TIIC. We over-plotted the PL/PW relations from Table \ref{tab_pl} and \ref{tab_pw}, together with the respected $\pm3\sigma$ boundaries, on the right panels of Figure \ref{fig_m31pc} after shifting these PL/PW relations vertically with $\mu=24.407$~mag \citep[][as black lines]{li2021}. Except for five TIIC that appeared to be brighter (marked as crosses in the right panels of Figure \ref{fig_m31pc}) in the $ri$-band PL relations, almost all of the TIIC were confined within the $\pm3\sigma$ of the respected PL/PW relations. Furthermore, scatters of these TIIC around the PL/PW relations confirmed the rather large dispersion in $ri$-band PL/PW relations as reported in Table \ref{tab_pl} and \ref{tab_pw}.

Our derived PL/PW relations can also be used to determine the distance modulus of M31 from this sample of TIIC (after excluding the five TIIC marked as crosses in the right panels of Figure \ref{fig_m31pc}). By fitting the data with the $ri$-band PL/PW relations given in Table \ref{tab_pl} and \ref{tab_pw}, weighted with the quadrature sums of errors on the mean magnitudes and the PL/PW dispersions, we obtained $\mu_r = 24.180\pm0.021$~mag, $\mu_i = 24.249\pm0.020$~mag, and $\mu_W = 24.423\pm0.026$~mag using the $ri$-band PL and PW relations, respectively. The quoted errors on $\mu$ are statistical errors only. The $\mu_W$ obtained from fitting the PW relation is in good agreements, and lie in between, the measurement of $\mu=24.407\pm0.032$~mag from \citet{li2021} and the recommended value of $\mu=24.46\pm0.10$~mag from \citet{dgb2014}. This suggested our derived $ri$-band PW relation is robust. On the other hand, distance moduli obtained from the $ri$-band PL relations are $\sim0.2$~mag smaller than $\mu_W$, hinting there could be additional systematic, in the order of $\sim0.2$~mag, in the derived PL relations. Distances to the globular clusters adopted from \citet{baumgardt2021} are unlikely to be the source of the systematic, because the same distances were used in deriving both of the PL and PW relations. Other possible systematic errors include the samples used, the extinction maps used, and the assumed extinction law to derive the $ri$-band PL relations. 

The derivation of $ri$-band PL relations include the TIIC sample transformed from the $BVI$-band photometry. Therefore, we first excluded the TIIC with transformations and only using the TIIC that have ZTF $ri$-band mean magnitudes, and re-derived the $ri$-band PL relations. Using the re-derived PL relations, the distance moduli of M31 we obtained are $\mu_r = 24.096\pm0.021$~mag and $\mu_i = 24.156\pm0.020$~mag. Similarly, we have used the ``SFD'' dust map for TIIC located outside the footprint of the {\tt Bayerstar2019} reddening map. If we re-derived the $ri$-band PL relations by using the same ``SFD'' dust map to all TIIC in the sample and re-determined the distance moduli to M31, then we obtained $\mu_r = 24.000\pm0.021$~mag and $\mu_i = 24.115\pm0.020$~mag. Finally, we adopted the same extinction law as in \citet{kodric2018}, i.e. $A_r = 2.554E$ and $A_i=1.893E$, and  we obtained $\mu_r = 24.150\pm0.021$~mag and $\mu_i = 24.229\pm0.020$~mag. These distance moduli are smaller than those obtained from the $ri$-band PL relations derived in Table \ref{tab_pl}. Hence, there could have hidden systematic errors when deriving the PL relations, and independent samples and calibration of the TIIC PL relations are desirable.

\section{Conclusions} \label{sec7}

In this work, we present the first $gri$-band and the updated $BVIJHK$-band PL and PW relations for TIIC located in the globular clusters. All-together, there are 70 TIIC spanning in 30 globular clusters (with ages spanning from $\sim11.0$ to $\sim13.2$~Gyr) in our sample, and only three of them have the complete nine band photometry. Homogeneous distance to the globular clusters, ranging from 3.30 (M22) to 88.47~kpc (NGC2419), adopted from a single source \citep{baumgardt2021} and consistent reddening maps, either the {\tt Bayerstar2019} 3D reddening map or the ``SFD'' dust map, were used to calibrate the absolute magnitudes of these samples of TIIC. We demonstrated that the PL relations are consistent in the $BVI$ and the $gri$ bands. We have also derived nine sets of the PW relations based on the combinations of these filters. For the PL/PW relations, the $JHK$-band PL/PW relations exhibit the smallest dispersion, which are preferable to be applied in the future distance scale work. Finally, our sample of TIIC also allow the derivation of PC and PQ relations in these filters. We found that the slopes of the PC relations in the $JHK$-band and in the $(r-i)$ color, as well as the slopes of the PQ relations, are quite shallow or flat. 

We tested our PL/PW relations, at least in the $ri$-band, with a sizable sample of TIIC in M31. The scatters of M31 TIIC on the PL/PW relations are similar to those presented in Table \ref{tab_pl} and \ref{tab_pw}, confirming the derived PL/PW dispersions are intrinsic. Using our derived $ri$-band PW relation, the distance modulus of M31 we obtained is in agreement with the latest measurement using the classical Cepheids. However, distance moduli derived from using the $ri$-band PL relations are smaller by $\sim0.2$~mag, suggesting there could be hidden systematics in the derived PL relations. Therefore, additional work in the near future are required to independently crosscheck these PL relations. Nevertheless, our derived PW relations can be applied in the on-going and upcoming synoptic time-series sky surveys, such as LSST or other surveys employing similar $gri$ filters.
 
\acknowledgments

We are thankful for the useful discussions and comments from an anonymous referee that improved the manuscript. We are thankful for funding from the Ministry of Science and Technology (Taiwan) under the contracts 107-2119-M-008-014-MY2, 107-2119-M-008-012, 108-2628-M-007-005-RSP and 109-2112-M-008-014-MY3. AB acknowledges funding from the European Union’s Horizon 2020 research and innovation programme under the Marie Skłodowska-Curie grant agreement No. 886298.

Based on observations obtained with the Samuel Oschin Telescope 48-inch Telescope at the Palomar Observatory as part of the Zwicky Transient Facility project. ZTF is supported by the National Science Foundation under Grants No. AST-1440341 and AST-2034437 and a collaboration including current partners Caltech, IPAC, the Weizmann Institute of Science, the Oskar Klein Center at Stockholm University, the University of Maryland, Deutsches Elektronen-Synchrotron and Humboldt University, the TANGO Consortium of Taiwan, the University of Wisconsin at Milwaukee, Trinity College Dublin, Lawrence Livermore National Laboratories, IN2P3, University of Warwick, Ruhr University Bochum, Northwestern University and former partners the University of Washington, Los Alamos National Laboratories, and Lawrence Berkeley National Laboratories. Operations are conducted by COO, IPAC, and UW.

This research has made use of the SIMBAD database and the VizieR catalogue access tool, operated at CDS, Strasbourg, France. This research made use of Astropy,\footnote{\url{http://www.astropy.org}} a community-developed core Python package for Astronomy \citep{astropy2013, astropy2018}.

\facility{PO:1.2m}

\software{{\tt astropy} \citep{astropy2013,astropy2018}, {\tt dustmaps} \citep{green2018}, {\tt gatspy} \citep{vdp2015}, {\tt Matplotlib} \citep{hunter2007},  {\tt NumPy} \citep{harris2020}, {\tt SciPy} \citep{virtanen2020}.}



\end{document}